\documentclass[12pt]{article}
\usepackage{authblk}
\usepackage{hyperref}
\hypersetup{
    colorlinks=true,
    linkcolor=black,     
    urlcolor=blue,
    citecolor=green,
    pdftitle={AdS/CFT}
    }
\usepackage[top=25truemm,bottom=28truemm,left=24truemm,right=24truemm]{geometry}
\usepackage{setspace}
\normalsize
\usepackage{amsmath,amssymb,mathrsfs,amsbsy,latexsym,amsfonts,amsthm}
\usepackage{fancyhdr}
\pdfoutput=1 
\usepackage{graphicx}
\usepackage{color}
\usepackage{comment}
\usepackage{here}
\usepackage{braket}
\usepackage[utf8]{inputenc}
\usepackage{mathtools}

\newcommand{\nn}{\nonumber\\}
\newcommand{\bO}{\mathbf{\Omega}}
\newcommand{\cH}{\mathcal{H}}

\DeclareMathOperator{\mir}{\mathsf{m}}
\DeclareMathOperator{\tr}{tr}
\usepackage{epsf}

\numberwithin{equation}{section}
\title{\bf Subregion Complementarity in AdS/CFT}

\author[1]{
Sotaro~Sugishita\thanks{\tt sotaro.sugishita(at)yukawa.kyoto-u.ac.jp}
}
\author[1]{
	Seiji~Terashima\thanks{\tt terasima(at)yukawa.kyoto-u.ac.jp}
}
\vspace{5mm}
\affil[1]{\it\normalsize 
Center for Gravitational Physics and Quantum Information,  
\mbox{Yukawa Institute for Theoretical Physics, Kyoto University, Kyoto 606-8502, Japan}  }
\date{}

\begin{document}

\maketitle
\thispagestyle{fancy}
\renewcommand{\headrulewidth}{0pt}

\begin{abstract}
We examine the bulk reconstruction in the AdS/CFT correspondence. We demonstrate that the subregion duality fails to hold, highlighting discrepancies between operators in causal wedge reconstruction and those in global reconstruction at the leading order in the large $N$ limit. 
We argue the invalidity of the entanglement wedge reconstruction based on the holographic quantum error correction code, attributing it to non-perturbative finite $N$ effects or quantum gravity effects due to the trans-Planckian modes near the horizon. Nevertheless, we propose the subregion complementarity, illustrating that different CFT operators can describe a bulk subregion. 
While we expect that this complementarity is valid outside the horizon in general eternal black holes, it is inapplicable for single-sided black holes where a semi-classical description at the stretched horizon is absent.
\end{abstract}

\newpage
\thispagestyle{empty}
\setcounter{tocdepth}{2}

\setlength{\abovedisplayskip}{12pt}
\setlength{\belowdisplayskip}{12pt}

\tableofcontents
\newpage
\section{Introduction and Summary}

The AdS/CFT correspondence \cite{Maldacena:1997re} claims that a certain conformal field theory (CFT) is dual to a quantum gravity with the asymptotic AdS boundary condition.
Assuming this correspondence, the CFT, called the holographic CFT, can be regarded as the definition of 
a quantum gravity.

The holographic CFT is typically realized by a large $N$ gauge theory with gauge group $SU(N)$.
Here, large $N$ means $N \gg 1$, but finite, otherwise it is not a well-defined quantum field theory.
On the bulk gravity side, $N$ is related to the ratio between the Planck length and the AdS length scale, and thus $N$ should be finite to keep both length scales finite.

The bulk semi-classical gravity description appears in the $1/N$ (asymptotic) expansion of this finite $N$ holographic CFT around the vacuum (or another semi-classical state).
In the leading order of the $1/N$ expansion around the global vacuum, we have the one-to-one correspondence between the low energy states of the holographic CFT and the free bulk states explicitly \cite{Fitzpatrick:2010zm, Terashima:2017gmc}.
In particular, for this case, the bulk local operator can be reconstructed from the CFT operator \cite{Hamilton:2006az} based 
on the earlier works \cite{Banks:1998dd, Bena:1999jv}. This is called the global reconstruction.

In a quantum gravity, it is important to consider a subregion of spacetime
because black hole physics is related to it.
In this context, the causal wedge reconstruction \cite{Hamilton:2006az} is anticipated to hold, where the bulk operators on the causal wedge corresponding to a CFT subregion $A$ can be reconstructed from the CFT operators on $A$.
It is also believed that this can be generalized to the entanglement wedge reconstruction \cite{Dong:2016eik} based on the agreement of the bulk and CFT relative entropy \cite{Jafferis:2015del}.
The correspondence between the reduced density matrices for bulk and CFT is also proposed \cite{Czech:2012bh, Bousso:2012sj} and is called the subregion duality.

In this paper, we examine the bulk reconstruction in the AdS/CFT correspondence for the subregion. 
We demonstrate that the subregion duality fails to hold.\footnote{
Here, the subregion duality and the entanglement wedge reconstruction means those given in \cite{Dong:2016eik} \cite{Almheiri:2014lwa}, in which the bulk Hilbert space was assumed to be a tensor product of the two Hilbert spaces for the subregion $M_A$ and $M_{\bar{A}}$.
This may be (approximately) realized by a gauge fixing,
for example, the FG gauge.
See section~\ref{subsec:contra}, especially eqs.~\eqref{sd} and \eqref{ewr}, 
for more precise statements of what we mean by the subregion duality and the entanglement wedge reconstruction. 
In particular, we claim that the global and Rindler HKLL bulk reconstructions of a bulk local operator in the overlap of the two entanglement wedges should be different at the leading correction of the $1/N$ expansion.
This contradicts the claim made in \cite{Almheiri:2014lwa} that a holographic quantum error-correcting code structure exists.
} 
Although the failure has been already claimed in \cite{Terashima:2020uqu, Terashima:2021klf, Sugishita:2022ldv, Terashima:2023mcr},
we give a definite example of the violations in this paper.
For clarification, we summarize our claim.
The global reconstruction says that for (smeared) bulk local operator $\phi$, we have a CFT operator $\phi^G$ acting on the entire space\footnote{More precisely, the support of $\phi^G$ is the CFT spacetime-region corresponding to the boundary region spacelikely separated from the bulk operator $\phi$.} where CFT lives.
The entanglement wedge reconstruction, in a strong sense, states that if $\phi$ is supported in the entanglement wedge of CFT subregion $A$, we have another CFT operator $\phi^R$ supported only on $A$ and the following equation holds
\begin{align}
    \phi^G \ket{\psi}=\phi^R\ket{\psi}
    \label{strong-ewr}
\end{align}
for arbitrary low-energy states $\ket{\psi}$.
In subsection \ref{subsec:contra} we consider the simplest case that the entanglement wedge of $A$ is the causal wedge, and explicitly show that discrepancies between $\phi^G$ and $\phi^R$ appear even at the leading perturbative correction in the $1/N$ expansion, and \eqref{strong-ewr} does not hold.

We here remark that we do not claim the invalidity of \eqref{strong-ewr} in the large $N$ limit.
Eq.\eqref{strong-ewr} can hold in the (UV-complete) bulk free theory, which corresponds to an appropriate $N=\infty$ limit of the holographic CFT (called generalized free theory). 
Since \eqref{strong-ewr} holds for $N=\infty$, one might think that even for large but finite $N$ one can reconstruct $\phi^R$ satisfying \eqref{strong-ewr} by adding $1/N$ corrections. 
Our claim is that this is impossible.
The counter-example of \eqref{strong-ewr} that we will show in subsection~\ref{subsec:contra} is based on the perturbative $1/N$ expansion. 
This implies that the bulk free theory (the $N=\infty$ generalized free theory) is essentially different from the theory with quantum gravity interactions (the theory with $1/N$ corrections).
Besides, in our previous paper \cite{Sugishita:2022ldv},  we argued that there is a problem connecting the AdS-Rindler quantization with the global quantization due to the cutoff of the semiclassical description in finite $N$ theory. 
In subsection~\ref{subsec:finiteN}, we argue that there is a potential issue in the entanglement wedge reconstruction due to non-perturbative finite $N$ effects or quantum gravity effects due to the trans-Planckian modes near the horizon following \cite{Sugishita:2022ldv}. 
This is related to the fact that the generalized free theory is not a good approximation of the finite $N$ CFT at high energy, and we have to be careful when we use the generalized free theory.
We note again that 
the discrepancies appear at
the (leading) perturbative correction in the $1/N$ expansion\footnote{Our claim is that the discrepancies between the causal wedge reconstruction and the global reconstruction cannot be resolved even if we take into account  the $1/N$ corrections of these reconstructions in the large $N$ expansion in the CFT side (or by appropriately solving the bulk interaction).}, as shown in subsection~\ref{subsec:contra},
although we emphasized that it is important to take $N$ finite, and the $1/N$ expansion will not be valid for some aspects. 
Here, the finite $N$ effect comes from the fact that the boundary of the subregion is strict and there is no notion of the $1/N$ corrected subregion.
Of course, this is assumed in the usual discussions of the entanglement wedges and then our claim is about the perturbative correction in the usual 
viewpoint.

It is important that 
the discrepancies between the global reconstruction and the Rindler reconstruction clearly 
contradict the holographic error correction code proposed in \cite{Almheiri:2014lwa}. This proposal implies that the Rindler reconstructions from different subregions should agree even with $1/N$ corrections, but, this agreement is explicitly limited to the low-energy (or code) subspace, addressing potential contradictions with axioms of quantum field theory. 
Our results demonstrate that such equivalence cannot hold even within the low-energy subspace.\footnote{As in [11], we define the code subspace as the low-energy states in the global quantization.
Instead, we may define another subspace as the low-energy space with respect to the AdS-Rindler Hamiltonian. However, the spectrum in AdS-Rindler quantization is continuous, while the total Hilbert space of finite $N$ CFT on the cylinder is discrete. Thus, we think that it is difficult to regard this low-energy space as the subspace of the total Hilbert space in finite $N$ although this disagreement may not be a problem for $N=\infty$.} (The global reconstruction can be viewed as a specific form of subregion reconstruction.)
Indeed, the quantum error correction structure is a key component of entanglement wedge reconstruction in \cite{Dong:2016eik}. This seems to rely on the assumption that, at least for low-energy states, the Hilbert space allows for factorization.

Our result seems to imply the AdS/CFT for the AdS-Rindler patch is not valid because 
the entanglement wedge reconstruction in \cite{Dong:2016eik} and the holographic error correction code proposal are not valid.
Nevertheless, we propose the subregion complementarity in section~\ref{sec:comp}, illustrating that different CFT operators can describe physics on a classical bulk subregion.
Indeed, the bulk local operators obtained by the AdS-Rindler HKLL bulk reconstruction \cite{Hamilton:2006az, Morrison:2014jha} gives
the correct correlation function in the AdS-Rindler patch in the large $N$ limit although they can be distinguished from the bulk local operators obtained by the global HKLL bulk reconstruction.
Thus, it serves as an alternative to the holographic error correction code.\footnote{The holographic error correction was introduced to resolve an apparent bulk locality paradox. 
This paradox arises when we suppose \eqref{strong-ewr} holds in the theory with the time-slice axiom.
Supposing $\phi^G$ and $\phi^R$ are different even in the low-energy subspace for finite $N$, there is no such paradox. 
In addition, the time-slice axiom does not hold in the generalized fee theory as an $N=\infty$ limit of the holographic CFT.
Therefore, we have no reason to introduce the holographic error correction structure for any $N$. 
}
In section~\ref{sec:bh}, we argue that this complementarity is valid outside the horizon in general eternal black holes. 
However, we discuss that in single-sided black holes, the complementarity cannot be applied because the semi-classical description is absent at the stretched horizon as expected from the brick-wall proposal \cite{tHooft:1984kcu, Iizuka:2013kma} and the fuzzball conjecture \cite{Mathur:2005zp, Mathur:2009hf}. 
It implies that the equivalence principle is violated near the horizon in single-sided black holes in quantum gravity.

To clarify the setup discussed here: we consider a connected subregion in an AdS vacuum background, representing the simplest configuration. We demonstrate that even in this case, the EWR in the sense of \cite{Dong:2016eik} is violated at order $1/N$, using only CFT-side arguments. This implies that the holographic quantum error correcting code (QECC) paradigm is fundamentally problematic at a basic level.
Note that in this paper, we focus on \cite{Dong:2016eik} and do not discuss
algebraic versions of the EWR discussed in \cite{Harlow:2016vwg} and \cite{Cotler:2017erl}.
As discussed in \cite{Sugishita:2024lee}, the description of the bulk operators is completely changed, and the holographic QECC structure is absent, same discussion in subsection \ref{subsec:contra} in this paper, 
for algebraic versions of the EWR.
Thus, for these, our result implies that the holographic QECC structure, which is an important and necessary ingredient of claims in \cite{Harlow:2016vwg} and \cite{Cotler:2017erl}, is absent.
At this point, we clarify the distinctions between our work and previous literature. To the best of our knowledge, no study other than ours provides a rigorous argument from the CFT perspective that the EWR assertion is violated at order $1/N$. While some works, such as \cite{Cotler:2017erl}, suggest that EWR might become approximate because the JLMS relation—including $1/N$ corrections—may not hold, our work is fundamentally different. In particular, since our results refute the existence of the holographic QECC, it is evident that our position stands in stark contrast to the prevailing consensus in the field.

\section{Remarks on gravitational theory on subregions}
\label{sec:remark}
\subsection{Difference between bulk semi-classical gravity and finite $N$ CFT}\label{subsec:finiteN}
In this subsection, we will summarize the results obtained in the previous paper \cite{Sugishita:2022ldv}, 
emphasizing the importance of the finite $N$ effects. 
In particular, we argue the importance of non-perturbative effects here, although leading perturbative $1/N$ corrections are relevant in the discussions in the next subsection.

\begin{figure}[htbp]
\centering
\includegraphics[width=4.5cm]{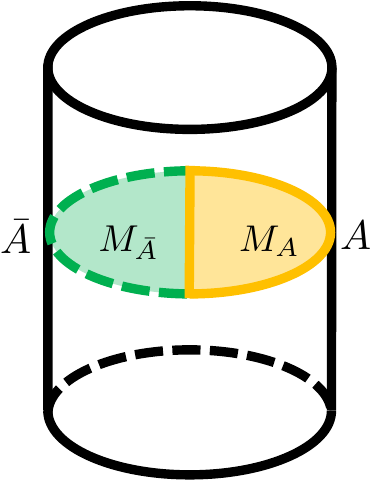}
\caption{Subregions. The yellow curve on the boundary represents a CFT subregion $A$ in a $\tau=\text{const.}$ slice, and the green dashed curve represents the complement subregion $\bar{A}$. $M_A$ is the bulk light yellow region associated with $A$. The bulk light green region is the complement subregion $M_{\bar{A}}$
}
\label{fig:A-MA}
\end{figure}

\begin{figure}[htbp]
\begin{center}
  \begin{minipage}[b]{.4\linewidth}
    \centering
    \includegraphics[height=1\linewidth]{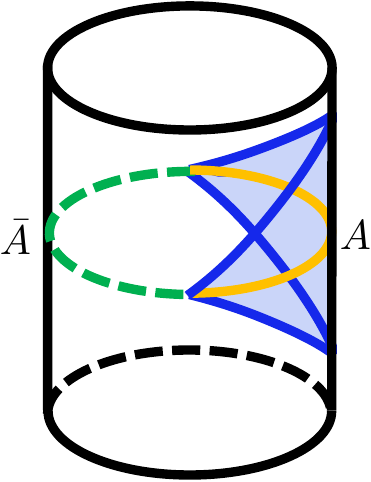}
  \end{minipage}%
  \begin{minipage}[b]{.4\linewidth}
    \centering
    \includegraphics[height=1\linewidth]{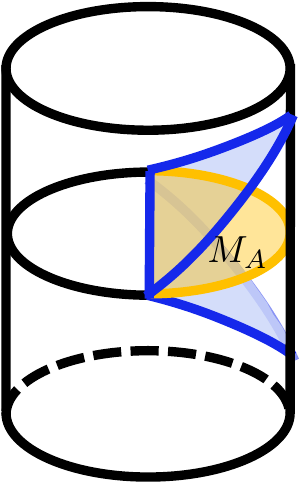}
  \end{minipage}\vspace{-1em}
  \end{center}
  \vspace{-1em}
  \caption{Causal domain $D(A)$ (left) and $D(M_A)$ (right). In the CFT side (the left figure), the causal domain $D(A)$ is the light blue region surrounded by the blue curves. In the bulk side (the right figure), the causal domain $D(M_A)$ is a spacetime subregion surrounded by the light blue surfaces and the cylinder.}
  \label{fig:D(A)}
\end{figure}

Here we first summarize the terminology of the AdS-Rindler wedge. More details of the convention of coordinates are summarized in Appendix~\ref{app:AdSRindler}. 
Let us consider the global AdS$_{d+1}$/CFT$_{d}$ correspondence where $\tau$ is the global time and the $\tau=$ const. surface in CFT is $S^{d-1}$.
We can take a spherical subregion $A$ in $S^{d-1}$ at $\tau=0$. The complement subregion is denoted by $\bar{A}$ (see Fig.~\ref{fig:A-MA}).
We can also consider the bulk subregion $M_A$ at bulk $\tau=0$ slice such that $M_A$ is enclosed by the asymptotic boundary corresponding to $A$ and the Ryu-Takayanagi surface \cite{Ryu:2006bv} associated with $A$. 
The bulk subregion $M_A$ is a time slice of the AdS-Rindler patch, i.e. the intersection of $\tau=0$ slice and the AdS-Rindler patch (see  Appendix \ref{app:AdSRindler} for the details of the AdS-Rindler patch). 
The complement bulk subregion is $M_{\bar{A}}$. Thus, we have $M=M_{A} \cup M_{\bar{A}}$ and $S^{d-1}=A \cup \bar{A}$, where 
$M$ is the bulk $\tau=0$ slice in the global AdS$_{d+1}$, 
and $S^{d-1}$ is the $\tau=0$ slice
of the cylinder ${\mathbf{R}}  \times S^{d-1}$ on which the CFT$_d$ lives.
We will denote the (union of future and past) domain of dependence, which is sometimes called the causal diamond, of $A$ as $D(A)$ and the one of $M_A$ as $D(M_A)$ which is the region that the AdS-Rindler patch covers, and we call $D(A)$ and $D(M_A)$ the Rindler wedge and the AdS-Rindler wedge respectively (see Fig.~\ref{fig:D(A)}).
In this case, the entanglement wedge is the same as the causal wedge, and we do not distinguish them.

In the global AdS$_{d+1}$/CFT$_{d}$ correspondence, $1/N$ expansion in the large $N$ limit of the holographic CFT corresponds to the perturbative expansion around the global (empty) AdS in the low energy. 
Our standard notion of bulk spacetime is only valid in this semi-classical approximation, although the true quantum gravity is defined by the finite $N$ CFT. 
Note that, in this paper, words like the gravity theory or the bulk theory mean this semi-classical theory. 
 
Since we have decomposed the bulk and CFT $\tau=0$ slices as $M=M_{A} \cup M_{\bar{A}}$ and $S^{d-1}=A \cup \bar{A}$, we can consider theories on subregions. 
The question is whether the semi-classical gravitational theory for the AdS-Rindler observer is dual to the CFT defined on the Rindler wedge.

\paragraph{Bulk AdS-Rindler} 
At the leading order of the semi-classical expansion, the bulk theory is a free theory on empty AdS$_{d+1}$ spacetime, and
the quantization of the free theory can be done.
In the AdS-Rindler wedge, we can take the AdS-Rindler time, and can take the creation/annihilation operators associated with this choice of time. 
As for the usual discussions on the Unruh effects, we can argue that the combined system of the free quantum theories on the AdS-Rindler patches $D(M_A)$ and  $D(M_{\bar{A}})$
is complete and there is the Bogoliubov transformation between 
the creation/annihilation operators of 
this system and those for the global time on the whole spacetime.\footnote{ 
The vacuum for the Rindler time is different from the global vacuum. 
}
In particular, the bulk local operators $\phi^R(X)$ expressed by the AdS-Rindler modes in $D(M_A)$ 
is identical to the ones $\phi(X)$ at the same points expressed by the global modes.
Here we put the superscript $R$ to the operators as $\phi^R(X)$ in order to distinguish the operators act on the bulk Hilbert space obtained by the AdS-Rindler quantization from that by the global time quantization although they are the same in the AdS-Rindler wedge in the (UV complete) bulk (free) theory.  For the global vacuum $\ket{0}$, we represent the reduced density matrix of the vacuum on $M_A$ by a thermal state $\sigma_{M_A}$, that is $\sigma_{M_A} =\tr_{M_{\bar{A}}} (\ket{0}\!\!\bra{0})$.

\paragraph{CFT Rindler} 
In the CFT picture, it is known \cite{Casini:2011kv} that the causal diamonds $D(A)$ of the spherical subregion $A$ in the whole spacetime $\mathbf{R}_\tau  \times S^{d-1}$ is conformal to $\mathbf{R}_{t_R} \times \mathbf{H}^{d-1}$ where $t_R$ is the Rindler time (the explicit map is given by \eqref{bd-glob-Rind} in Appendix~\ref{app:AdSRindler}).
Thus, the CFT on the Rindler wedge $D(A)$ is just the CFT on $\mathbf{R} \times \mathbf{H}^{d-1}$. 
In particular, for $d=2$ it is the CFT on the Minkowski spacetime $\mathbf{R}^{1,1}$.

\paragraph{A paradox in AdS/CFT} 
The above bulk and CFT descriptions can be considered independently and both are correct.
In AdS/CFT, the bulk picture is equivalent to the low energy (and large $N$) approximation of CFT.
The operators in these two descriptions are related, by the BDHM relation \cite{Banks:1998dd}, which claims that the bulk field $\phi(X)$ becomes the CFT primary field $\mathcal{O}^\text{CFT} (x) $ by bringing it to the boundary with an appropriate scaling factor.  
We assume the BDHM relation for the global AdS/CFT correspondence. We now consider the bulk free field $\phi^R(X)$ on the bulk subregion $M_A$. 
We will bring it to the boundary and denote it by $\mathcal{O} (x) $ with an appropriate factor.
Note that this bulk field $\phi^R(X)$ is equivalent to the bulk field $\phi(X)$ defined in the global AdS as stated above. 
Thus, according to the BDHM relation,
$\mathcal{O} (x) $ should become the CFT primary field $\mathcal{O}^\text{CFT} (x)$ on $A$ (up to a scaling factor). In particular for $d=2$, the CFT on $A$ is that on 
Minkowski spacetime.
However, the  $\mathcal{O} (x) $ cannot be $\mathcal{O}^\text{CFT} (x)$ for finite $N$ because $\mathcal{O} (x) $ contains the creation operators associated with modes $e^{-i\omega t + i p x}$ with $\omega^2-p^2<0$ and such tachyonic operators\footnote{
In the bulk theory in the AdS-Rindler wedge, these operators are not tachyonic. 
} cannot exist.
Note that the energy of the tachyonic creation operators can be negative by the Lorentz transformation.
This seems to be a paradox \cite{Sugishita:2022ldv}.

Here we remark on the state-dependence of the bulk descriptions. 
That is, the bulk backgrounds depend on what CFT states we consider.
If we choose a state that corresponds to a semi-classical bulk background, like the pure AdS or the AdS-Schwarzschild,
we will have a free theory on the background and 
the bulk reconstruction of CFT operators will depend on the state.
In the CFT side of  
the paradox we are discussing,
there seem to appear two semi-classical backgrounds.
When we consider the global vacuum, the dual background is the global AdS.
The CFT reduced density matrix on the subregion $A$ for this global vacuum is the thermal state with respect to the Rindler Hamiltonian. 
The dual background for this mixed state is the black hole background and it is equivalent to the AdS-Rindler background (we will review this fact in subsection~\ref{sec:Poincare}).
On the other hand, we can also consider the vacuum state for the Rindler Hamiltonian (Rindler vacuum). 
As we said, the CFT on $D(A)$ is just the  $\mathbf{R} \times \mathbf{H}^{d-1}$. In particular, for $d=2$, it is CFT on $\mathbf{R}^{1,1}$, and thus the Rindler vacuum is the Poincare vacuum for this CFT. 
The dual bulk background for the Poincare vacuum is the Poincare AdS spacetime. 
If we consider the bulk (free) theory on the Poincare AdS geometry, there are no tachyonic modes, unlike the AdS-Rindler case. 

On the other hand, on the bulk theory side, 
only the pure AdS background appears.
Indeed, the global bulk reconstruction is based on the global vacuum of the AdS background. 
The AdS-Rindler bulk reconstruction is based on the Rindler vacuum of the AdS background and the global vacuum is the thermal excitation from it on the same semi-classical background.
Thus, it assumes the spectrum including the tachyonic modes around the Rindler vacuum, 
which is different from the one for the (finite $N$) or the Poincare AdS.
This may imply that the large $N$ limit and taking subregion do not commute with each other.\footnote{In the above paradox, on the CFT side, we first have the finite $N$ holographic CFT, consider a subregion (the Rindler wedge). 
Then, the thermal state for the Rindler Hamiltonian does not contain the tachyonic modes, and it thus dose even after we take the large $N$ limit. 
On the bulk side, we first start from the bulk free theory and then consider the AdS-Rindler wedge. 
We thus take the large $N$ limit first in the bulk side. Indeed, the bulk subregion is a vague concept in finite $N$ unlike the CFT side, because the spacetime notion makes sense only after taking the classical limit. 
The order of the large $N$ and taking the subregion is different for the bulk and CFT in the above paradox.}  

However, we should take care that the spectra are obtained by using the bulk description which is just the effective description of the true finite $N$ CFT. 
For the finite $N$ CFT, we expect that the Hilbert space obtained by the Rindler time quantization of the CFT is complete on $D(A)$ such that any reduced density matrices on $A$ obtained from global states can be represented as density matrices in this Rindler Hilbert space%
.\footnote{The algebra on the subregion $D(A)$ is type III and we cannot define the reduced density matrix in a rigorous sense, unless we do not introduce some cutoff. We here assume that we have a cutoff such that the algebra on $D(A)$ is type I.
Note that the auxiliary UV cutoff in this CFT, $\Lambda_{CFT}$, can be taken to be arbitrarily large; in particular, it can be chosen to be significantly larger than the Planck mass, $\Lambda_{pl}$, which depends on $N$. While $\Lambda_{pl}$ is the breakdown scale of the low-energy bulk effective theory, the auxiliary $\Lambda_{CFT}$ is directly unrelated to it.
}

\paragraph{Origin of the paradox} 
Here, it is important to note that the bulk free theory, which is the generalized free theory in the CFT picture, is only the low energy and large $N$ approximation of the (finite $N$) CFT.
The origin of the above paradox is that we trust the approximation beyond the extent to which it is valid.
When we consider the CFT with the global vacuum state, the bulk description is valid only for the low-energy states with respect to the global Hamiltonian. 
Then, the bulk free theory on the AdS-Rindler patch $D(M_A)$ is a good description for the CFT on $D(A)$ only for the low-energy sectors of the free theory on the global patch.
However, some low-energy modes in the AdS-Rindler time quantization contain arbitrary high-energy modes for the global time quantization \cite{Sugishita:2022ldv}.\footnote{This is not a problem if we are interested in the UV complete bulk free theory. This UV complete theory is the generalized free theory in the CFT picture. The approximation of the finite $N$ CFT by the generalized free theory is valid only at low-energy (and large $N$).}
Thus, due to these problematic modes, the free theory on the AdS-Rindler patch is not always valid as an approximation of the finite $N$ CFT. 
There is no paradox if we take care of the effectiveness of the bulk description. 

The problematic modes are indeed related to the above tachyonic modes and also the geodesics connecting to the past and future horizon \cite{Sugishita:2022ldv}. Essentially, these modes are dominant around the horizon as seen from the suppression of the eigenfunction for the radial direction, and they are suppressed if we consider the smeared bulk operators supported on subregions in $D(M_A)$ apart from the horizon. 
Thus, in spite of the above problem, the bulk free description is valid except for the stretched horizon region, and thus the following equation, which exactly holds in the free theory, 
\begin{align}
   \bra{0} \phi(X) \phi(X') \ket{0} 
   =\tr_{M_A} 
   (\sigma_{M_A} \phi^R (X) \phi^R (X') ) ,
\end{align} 
approximately holds even for the gravity dual of the finite $N$ holographic CFT, if the bulk operators are smeared appropriately. 
This is the subject we will consider in section~\ref{sec:comp}.

A general lesson here from the above discussion is that the bulk semi-classical gravitational theory as the low-energy effective theory of quantum gravity  (= the finite $N$ CFT) is observer-dependent. 
Specifically, its validity depends on the chosen patch and its associated time foliation. When we consider a causal diamond of a spacetime subregion, like the AdS-Rindler wedge $D(M_A)$, we encounter ``horizons'' corresponding to the boundary of the causal diamond. 
A semi-classical gravitational theory formulated within these causal diamonds, accompanied by a time foliation that causes an innumerable number of time slices to converge near the horizon (akin to the Rindler time), becomes unreliable near the horizon as the approximation of the finite $N$ CFT. 
This unreliability is attributed to the UV cut-off, typically associated with the Planck mass, of the bulk effective theory. 
We stress that the breakdown of the effective theory can be seen by examining the finite $N$ effects.
This is because the $1/N$ expansion (i.e. the semi-classical expansion) is based on the leading-order spectrum which is obtained by taking $N=\infty$,  
and the perturbative corrections do not change the leading spectrum.
In this sense, we need the non-perturbative quantum gravity effect. 
The non-commutativity between the large $N$ limit and taking a subregion argued above might also come from ignoring the non-perturbative effect. 
The true theory is quantum gravity defined by the finite $N$ CFT, and the semi-classical gravity obtained by the perturbative $1/N$ expansion around $N=\infty$ is just an effective description. 
For large but finite $N$, we need non-perturbative effects which especially cannot be negligible near the ``horizon''  for an observer in the subregion.

We call the semi-classical theory for the observer in a subregion (i.e. theory with the patch and time-foliation inside a causal diamond) the bulk theory on the subregion.
The statement that the bulk theory on the subregion is not always applicable might be surprising. 
This is because the semi-classical expansion of the bulk gravity theory is anticipated to remain valid even within spacetime subregions, such as the Rindler region or the more general causal diamonds. 
Such an approach has been employed extensively in numerous studies. 
For instance, the AdS-Rindler wedge reconstruction—a specific case of the entanglement wedge construction—relies on this very assumption. 
In the following subsection, we will demonstrate how the subregion duality and the entanglement wedge reconstruction do not work. We here remark that the subregion duality may hold between the bulk free theory and the generalized free theory (or extension of them including $1/N$ perturbative corrections). 
Nevertheless, the generalized free theory is not a good approximation of a finite $N$ CFT at high energy and we must take account of the non-perturbative corrections.

On the other hand, the invalidity of the semi-classical description near the horizon (or at the stretched horizon) is known as the trans-Planckian problem \cite{tHooft:1984kcu}. This is due to the use of the coordinates which are singular at the horizon. 
It may imply that UV physics is important near the horizon and cannot be neglected as suggested by the brick wall model \cite{tHooft:1984kcu} or the fuzzball conjecture \cite{Mathur:2005zp, Mathur:2009hf}. 
Here, we note that in order to discuss the bulk reconstruction or the entanglement entropy for a subregion,
the full (entanglement) spectrum of the subregion is needed. It means that we need the whole subregion including the vicinity of the horizon.\footnote{
If the smeared bulk local operator in the AdS-Rindler patch is not close to the horizon, 
it contains almost the low energy modes only. Here, the smearing is defined by essentially the Gaussian in the AdS-Rindler coordinate (See Appendix \ref{mode}).
}

\subsection{Failure of the subregion duality and entanglement wedge reconstruction} 
\label{subsec:contra}

In the previous subsection, we have argued that it is dangerous to trust the semi-classical bulk theory when we consider a subregion. 
Indeed, the subregion duality, which might be expected to hold from the bulk semi-classical point of view, does not hold in finite $N$ CFT. 
In this subsection, we explain in what sense the subregion duality and the causal (and also entanglement) wedge reconstruction do not work.

The subregion duality  is the claim that for any states $\rho, \sigma$ (around the global vacuum), we have 
\begin{align}
   \rho_A =\sigma_A \Longleftrightarrow 
   \rho_{M_A}=\sigma_{M_A},
   \label{sd}
\end{align}
for a subregion $A$ for the CFT and the bulk subregion $M_A$, which is a time-slice of the entanglement wedge of $A$, for the bulk theory where $\rho_A =\tr_{\bar{A}} (\rho)$ and 
$\rho_{M_A} =\tr_{M_{\bar{A}}} (\rho)$.
Note that the low-energy Hilbert space $\cH$ is the same in CFT and the bulk theory, and $\rho, \sigma$ are the density matrices in this space. $\cH$ is formally decomposed into $\cH_A \otimes \cH_{\bar{A}}$ in the CFT side, but this decomposition is in principle not directly related to the decomposition in the bulk side $\cH_{M_A} \otimes \cH_{M_{\bar{A}}}$.

Before proceeding, we have to remark on the problem of the gauge fixing in the bulk theory.
In the above, we have assumed that the bulk Hilbert space is factorized as $\cH_{M_A} \otimes \cH_{M_{\bar{A}}}$ in order to define the bulk reduced density matrices $\rho_{M_A}, \sigma_{M_A}$. This is a standard assumption when we discuss the subregion duality or the entanglement wedge reconstruction \cite{Dong:2016eik}. 
However, the notion of the locality is ambiguous in the gravitational theory due to the diffeomorphism. 
For example, the bulk scalar field $\phi(x)$ is not gauge-invariant\footnote{\label{foot_weak}To have a gauge invariant quantity, we have to attach a gravitational dressing to $\phi(x)$ like the Wilson line in the gauge theory \cite{Donnelly:2015hta, Donnelly:2016rvo, Giddings:2018umg, Giddings:2019hjc}. We can define operators supported on the bulk subregion $M_A$ as operators whose gravitational dressing is also supported only on $M_A$. For these restricted operators, the subregion duality may hold, which is the same as we call the weak version of the subregion duality in \cite{Terashima:2021klf, Sugishita:2022ldv}. 
The JLMS formula \eqref{eq:jlms} in the sense of this weak version may also be valid. However, this weaker version does not assume the holographic QECC structure.} 
and does not make sense unless we fix the gauge (coordinates system). 
Thus, the notion of $\cH_{M_A}$ and $\cH_{M_{\bar{A}}}$ makes sense only after the gauge fixing of the bulk diffeomorphism. 
Note that the HKLL bulk reconstructions are indeed done in the FG gauge.
In this sense, the above $\rho_{M_A}, \sigma_{M_A}$ is not gauge invariant and may depend on a choice of the gauge, for example, the gauge used in the global HKLL bulk reconstruction.
Our aim in this subsection is to show the subregion duality is invalid under the standard assumption of the factorization $\cH_{M_A} \otimes \cH_{M_{\bar{A}}}$. Thus, we proceed with our discussion by assuming that a gauge is fixed. 
To discuss it in a gauge invariant way, we need an algebraic approach as we do in the definition of entanglement entropy in gauge theories. We leave it to future works.

The claim \eqref{sd} is equivalent to that the relative entropies are the same, under the assumption of the factorization $\cH_{M_A} \otimes \cH_{M_{\bar{A}}}$:
\begin{align}
\label{eq:jlms}
   S(\rho_A |\sigma_A) =
   S(\rho_{M_A} | \sigma_{M_A}),
\end{align}
which was shown in \cite{Jafferis:2015del} by using the semi-classical bulk computations
\cite{Faulkner:2013ana}.
The entanglement wedge reconstruction
straightforwardly follows from the subregion duality \eqref{sd} \cite{Dong:2016eik}.
It claims that for any (low-energy) bulk operator $\phi_{M_A}$ supported in $M_A$
there exists the CFT operator $\mathcal{O}_A$ supported in $A$ such that
\begin{align}
   \phi_{M_A} \ket{\psi} = \mathcal{O}_A \ket{\psi},
   \label{ewr}
\end{align}
in the low energy approximation,\footnote{More precisely, the equality in \eqref{eq:jlms} and \eqref{ewr} should be understood as the approximate equality $\simeq$ in the low energy and large $N$ approximation. We will use $=$ instead of $\simeq$ just for simplicity.} where $\ket{\psi}$ are any states in the low energy subspace around the global vacuum $\ket{0}$.
Via the global bulk reconstruction \cite{Hamilton:2006az}, $\phi_{M_A}$ is a CFT operator supported on the entire space $S^{d-1}$. 
Thus, \eqref{ewr} claims that there exists a CFT operator $\mathcal{O}_A$ supported on subregion $A$ for a CFT operator $\mathcal{O}$ supported on the whole space, such that $\mathcal{O}_A$  acts on $\ket{\psi}$ as $\mathcal{O}$ does. 
Here, it is important that the reconstructed operator $\mathcal{O}_A$ does not depend on $\ket{\psi}$  in the low-energy subspace, like the HKLL reconstruction.
This also implies that 
if $\mathcal{O}_A^1, \mathcal{O}_A^2$ are CFT operators corresponding to bulk low-energy operators $\phi_{M_A}^1, \phi_{M_A}^2$, the CFT operator corresponding to the product operator $\phi_{M_A}^1\phi_{M_A}^2$ must be $\mathcal{O}_A^1 \mathcal{O}_A^2$
because $\ket{\psi'} \equiv \phi_{M_A}^2 \ket{\psi}= \mathcal{O}_A^2 \ket{\psi}$ 
is also in the low-energy subspace.
Thus, this is the operator reconstruction (in the subspace) and not the state reconstruction.
The reconstruction of the state is rather easy.
Indeed, via the Reeh–Schlieder theorem, we can reconstruct arbitrary states from operators in any subregion, not limited to the entanglement wedge, by acting them on the vacuum. 
In other words, if one interprets entanglement wedge reconstruction just as the reconstruction of states, one no longer needs to insist on the entanglement wedge.
The original entanglement wedge reconstruction claim should be interpreted as a statement about operator reconstruction, not the state reconstruction.

It is important to note that the ``derivation'' of the subregion duality and entanglement wedge reconstruction is based on the semi-classical expansion. 
The leading part of the expansion is described by the generalized free theory. 
As we stated above, the generalized free theory is different from the finite $N$ CFT at high energy. 
Entanglement entropy is a UV-dependent quantity, and thus there is no reason that the entropy in the generalized free theory is the same as that in the finite $N$ CFT, even if we compare the universal terms of the entropy. 
For instance in \cite{Faulkner:2013ana}, the bulk entanglement entropy is computed based on the semi-classical expansion where the 1-loop computations are based on the generalized free approximation.  
The bulk modular Hamiltonian (for bulk free fields) is the modular Hamiltonian for the generalized free on the boundary theory which is not that for the finite $N$ CFT. 
Indeed, the bulk modular Hamiltonian (which is the AdS-Rindler Hamiltonian for the AdS-Rindler case) contains arbitrary high energy modes for the global Hamiltonian \cite{Sugishita:2022ldv} and we cannot trust it because it is beyond the low-energy approximation. 
To compute the bulk entanglement entropy, we need information at the short distance where we expect the non-perturbative quantum gravity effects to be non-negligible.
As stressed above, the bulk theory on a subregion cannot be regarded as the low-energy approximation of the CFT for the region near the horizon (entangling surface caused by taking the subregion), while the vicinity of the entangling surface dominantly contributes to the entanglement entropy.
Thus, the discussion based on the bulk entanglement (i.e., entanglement for generalized free theory) has subtleties and might not be reliable for the approximation of the finite $N$ CFT, at least for the corrections to the leading order in the $1/N$ expansion.

When we treat the semi-classical bulk theory as a UV complete theory, the subregion duality and entanglement wedge reconstruction are applicable (see e.g. \cite{Leutheusser:2021frk, Leutheusser:2022bgi}), that is, they hold, at least for the smeared low-energy operators not close to the horizon of the causal diamond, in the generalized free theory, not the full CFT, although high-energy modes appear near the horizon (see Appendix~\ref{mode}). 
On the other hand, when we view the bulk theory as the low-energy approximation of the CFT, we encounter issues with both conformal symmetry and unitarity. 
Such contradictions were already found as the radial locality paradox in \cite{Almheiri:2014lwa}, and it was assumed to be resolved by the existence of the quantum error correction code-like structure. However, we assert that such a structure is non-existent because all paradoxes come from the inappropriate use of the generalized free description.

\begin{figure}[htbp]
\centering
\includegraphics[width=8cm]{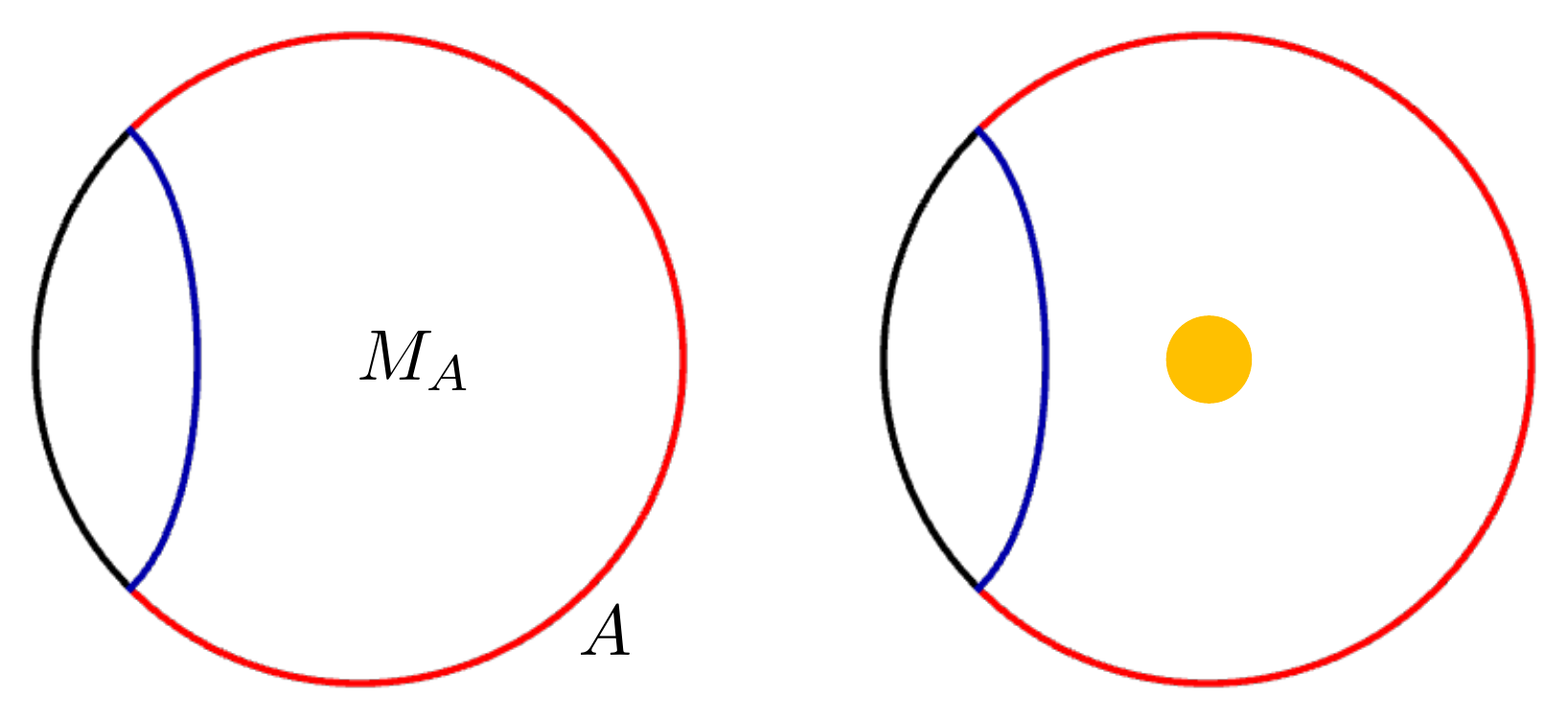}
\caption{AdS-Rinlder subregion. In the left figure, $M_A$ is the bulk subregion enclosed by the boundary subregion $A$ (red curve) and the Ryu-Takayanagi surface (blue curve). The smearing function $K$ has a support (orange region in the right figure) which is spherically symmetric and included in $M_A$.
}
\label{fig:smear}
\end{figure}

Below, we will explicitly demonstrate the invalidity of the subregion duality and the entanglement wedge reconstruction, by showing a case that both sides of eq.~\eqref{ewr} are different even at the leading order correction of the large $N$ expansion.\footnote{
To show it, we will use the micro causality of the boundary theory which will be valid for the finite $N$ CFT.
}
We here take a spherical subregion $A$ for the CFT defined on the sphere as in the previous subsection. Then, the entanglement wedge (causal wedge) in the bulk is the AdS-Rindler patch in the global AdS, i.e., the associated bulk subregion is $M_A$ defined in the previous subsection. 
We now consider smeared bulk local operators so that our discussion is closed in the low-energy sector as
\begin{align}
   \tilde{\phi}_{M_A} =\int dX' K(X') \phi(X'),
   \label{ew00}
\end{align}
where $K(X')$ is a smearing function (or distribution) supported only on bulk subregion $M_A$. Do not confuse this smearing with smearing used in the HKLL reconstruction. 
Here the smearing function $K$ in \eqref{ew00} is introduced just to consider a low-energy operator and is arbitrary if it is a smooth function and its support is large enough such that it contains mostly low energy modes. 
Via the global HKLL reconstruction \cite{Hamilton:2006az}, $\phi(X')$ is a Hermitian CFT operator supported on the entire region $S^{d-1}$, and thus $\tilde{\phi}_{M_A}$ is a smooth function such that it contains mostly low energy modes.
We consider the following state given by the unitary operation:
\begin{align}
   \ket{K} = e^{i \epsilon \tilde{\phi}_{M_A}} \ket{0}, 
   \label{ewu0}
\end{align}
where the smearing function $K$ was chosen to be real and $\epsilon$ is a real small parameter and thus $e^{i \epsilon \tilde{\phi}_{M_A}}$ is unitary. 
Let $\rho, \sigma$ be the density matrix for $\ket{K}$ and $\ket{0}$ as $\rho=\ket{K}\bra{K}$, $\sigma=\ket{0}\bra{0}$.
Note that we cannot distinguish $\ket{K}$ with the global vacuum in $M_{\bar{A}}$, i.e. 
$\rho_{M_{\bar{A}}}= \sigma_{M_{\bar{A}}}$  where $ \rho_{M_{\bar{A}}} \equiv \tr_{M_A} \rho,  \sigma_{M_{\bar{A}}} \equiv \tr_{M_A} \sigma$ because $e^{i \epsilon \tilde{\phi}_{M_A}}$ is a unitary operator acting only on $\cH_{M_A}$.
Let us take the CFT subregion $A$ bigger than the half of the whole space $S^{d-1}$ and consider the smearing function spherically symmetric around the center of AdS space and supported on a small region such that the smearing function $K$ is supported on $M_A$ although it is much larger than the cut-off length (see Figure~\ref{fig:smear}).\footnote{
If we want to take $A$ as the half space of the entire sphere, we can move the bulk local operator slightly toward $M_A$ by the conformal transformation such that
the bulk operator $\phi_{M_A}$ is supported on $M_A$.
It should have a non-zero energy density for any point, in the CFT picture, by the continuity.
The discussions below can be valid for this modification.
}
Then, it should have a spherically symmetric energy density in the CFT picture.
Using the global HKLL bulk reconstruction, this state is given by
\begin{align}
   \ket{K} = e^{i \epsilon \int d \tau \tilde{K}(\tau) \int d\Omega \, \mathcal{O}(\tau, \Omega) } \ket{0},
   \label{ewu1}
\end{align}
where $\tilde{K}(\tau)$ is a function of $\tau$ depending on $K$, but $\Omega$-integration is spherically symmetric.  
Then, we find $  \rho_{\bar{A}}  \neq \sigma_{\bar{A}}$ because 
the unitary operator acts on the entire region including $\bar{A}$ for $\ket{K}$ in \eqref{ewu1}.
Thus, the subregion duality \eqref{sd} is
violated where $\sigma=\ket{0} \bra{0}$.
(The roles of $A$ and $\bar{A}$ are reversed.)\footnote{
We again remark that this conclusion $\rho_{M_{\bar{A}}}= \sigma_{M_{\bar{A}}}$ but $\rho_{\bar{A}}\neq  \sigma_{\bar{A}}$ relies on the factorization $\cH_{M_A} \otimes \cH_{M_{\bar{A}}}$ with a gauge-fixing as remarked in the beginning of this subsection. 
The assumption $\cH_{M_A} \otimes \cH_{M_{\bar{A}}}$ enables us to define bulk reduced density matrices $\rho_{M_A}, \sigma_{M_A}$ in \eqref{sd}.
If we take a gauge-invariant approach, the above $\tilde{\phi}_{M_A}$ should have a gravitational dress acting on $M_{\bar{A}}$ (see also \cite{Sugishita:2024lee}), and thus the above discussion is not valid. Taking an alternative definition of reduced density matrices, we have a weak version of the subregion duality as noted in footnote~\ref{foot_weak}.
}

The entanglement wedge reconstruction \eqref{ewr}, which is now the causal wedge reconstruction, requires that there exists the state 
\begin{align}
   \ket{K^R}=e^{i \epsilon \tilde{\phi}^R_{M_A}} \ket{0},
   \label{ewu01}
\end{align}
such that $\tilde{\phi}^R_{M_A}$ is constructed from the CFT operators only supported on $A$ such that $\ket{K}=\ket{K^R}$.\footnote{
Here, we note again that the bulk operator $\tilde{\phi}$, not just the state $\tilde{\phi}^n \ket{0}$, should be reconstructed.
Indeed, the global and AdS-Rindler HKLL bulk reconstructions give such an operator
and we will see later that the reconstructed CFT operators
give correct bulk correlation functions
even though the subregion duality and the entanglement wedge reconstruction are not valid. 
If we just want to reconstruct the states, the CFT operators on $A$, for the Rindler wedge case, can give any CFT (and bulk) state and the notion of entanglement wedge reconstruction is useless.
This is because 
using the mirror operator for the thermofield double state (see Appendix~\ref{App:mirror}),
we can obtain $\mathcal{O}_{\bar{A}} \ket{0}$, where $\mathcal{O}_{\bar{A}}$ is any CFT operator supported on $\bar{A}$ by
$\mathcal{O}_{A} \ket{0}$, where $\mathcal{O}_{A}$ is a CFT operator supported on $A$.
More generally, via the Reeh–Schlieder theorem, we can reconstruct arbitrary states from any subregion not limited to the entanglement wedge.
}
Let us confirm that there exist no such operators $\tilde{\phi}^R_{M_A}$ by comparing the energy density $ T_{00}(\tau=0, x)$ at $x$ for the two states $\ket{K}$ and $\ket{K^R}$ in the CFT picture.
Indeed, generically,
\begin{align}
    \bra{K } \,
    T_{00}(\tau=0, x) \, \ket{K} 
    \neq  \bra{K^R } \,
    T_{00}(\tau=0, x) \, \ket{K^R}  ,
\end{align}
for the boundary point $x$ space-like separated from all points of the causal diamonds of $A$,
for example, $x$ in $\bar{A}$ (at $\tau=0$).
Because of the spherical symmetry of $\ket{K}$, it is clear that 
\begin{align}
    \bra{K } \,
    T_{00}(\tau=0, x) \, \ket{K}
    = \frac{1}{\text{Vol}(S^d)} \bra{K } \,
     H \, \ket{K} =\frac{1}{\text{Vol}(S^d)} (\epsilon^2 \bra{0 } \tilde{\phi}_{M_A}\,
    H\, \tilde{\phi}_{M_A} \ket{0}+ \mathcal{O}(\epsilon^3)),
\end{align}
where $H$ is the Hamiltonian,
is non-zero and $\mathcal{O}(\epsilon^2)$.
Here, we subtracted the Casimir energy from the definition of $T_{00}$ and the Hamiltonian such that $\bra{0} T_{00}(\tau=0, x) \ket{0}=0$.
On the other hand, 
\begin{align}
 & \bra{K^R } \,
    T_{00}(\tau=0, x) \, \ket{K^R} \nn
    &=\bra{0} (T_{00}(\tau=0, x) + [-i \epsilon \tilde{\phi}^R_{M_A}  ,
    T_{00}(\tau=0, x) ]+  \cdots)\ket{0}
    =0,
\end{align}
exactly by the causality of CFT.\footnote{
Note that without changing the discussion here, we can smear $T_{00}(\tau=0, x)$ appropriately in order to consider the low energy modes only. Thus, even in the low energy approximation, \eqref{ew1} does not hold.
} 
Thus, $\ket{K}$ and $\ket{K^R}$ cannot be the same states and thus the entanglement wedge reconstruction \eqref{ewr} does not hold.  
Note that we have used only the three-point functions (single $T$ and two $\mathcal{O}$) for $\mathcal{O}(\epsilon^2)$. 
Therefore, \eqref{ewr} is invalid even at the leading order in the large $N$ expansion.

The arguments presented so far remain valid even if we consider the operator $T_{00}$ smeared over the (domain of dependence of) $\bar{A}$, since only the linearity of $T_{00}$ has been exploited. In particular, by smearing, we can transform it into a low-energy operator, thereby restricting the discussion to the subspace of low-energy states.

We can repeat this discussion for truncated states $\ket{K} = \sum_{n=0}^{n_0} \frac{1}{n!} (i \epsilon \tilde{\phi}_{M_A})^n \ket{0}$ and $\ket{K^R} = \sum_{n=0}^{n_0} \frac{1}{n!} (i \epsilon \tilde{\phi}^R_{M_A})^n \ket{0}$, which are the polynomials of the bulk local fields.
Then, the energy density for $\ket{K}$ is uniform and approximately $ \bra{0} (\epsilon \tilde{\phi}_{M_A})  T_{00} ( \epsilon \tilde{\phi}_{M_A}) \ket{0}$ which is 
$\mathcal{O}(\epsilon^2)$.
On the other hand, by the causality of CFT we can show 
$\bra{K^R } \,
    T_{00}(\tau=0, x) \, \ket{K^R}  
    =\bra{0} (T_{00}(\tau=0, x) + [-i \epsilon \tilde{\phi}^R_{M_A}  ,
    T_{00}(\tau=0, x) ]+  \cdots)\ket{0}
    =\mathcal{O}(\epsilon^{n_0+1}) $,
thus $\bra{K } \,
    T_{00}(\tau=0, x) \, \ket{K}  \neq \bra{K^R } \,
    T_{00}(\tau=0, x) \, \ket{K^R} $.
Therefore, $\ket{K^R}$ cannot be the same as $\ket{K}$, and $\tilde{\phi}_{M_A}$ and $\tilde{\phi}^R_{M_A}$ are completely different operators.

We have explicitly seen that the global and AdS-Rindler HKLL bulk reconstructions are different.
Let us consider why they are different.\footnote{
The bulk local operator $\phi(X)$ is defined in a gauge choice and these two bulk reconstructions may correspond to different gauge choices because the radial coordinates $\rho$ and $\xi$ are different.
This may be a reason for the difference.
Below, we will explain why they are different in
the $1/N$ expansion from the $N=\infty$ theory, where there is no difference. 
} 
If we regard the bulk free theory as the UV complete theory,
\begin{align}
   \bra{0} \phi^G (X) \phi^G (X') \ket{0} 
   =\tr_{M_A} 
   (\sigma_{M_A} \phi^R (X) \phi^R (X') ) ,
   \label{eq311}
\end{align}
where $\phi^R (X)$ is constructed from the mode expansion in the right AdS-Ridnler wedge, 
will hold as we can see from the usual discussion on the Unruh effect in Minkowski space.
If $X,X'$ are not close to the AdS-Rindler horizon, \eqref{eq311} approximately holds for the gravity dual of the finite $N$ holographic CFT as explained in section~\ref{sec:comp}.
However, it does not hold if  $X,X'$ are close to the AdS-Rindler horizon (see Appendix \ref{mode}).
The HKLL bulk reconstructions use the equations of motion of the bulk theory in order to relate the bulk local field and CFT operators.
However, the equations of motion in the bulk low energy description are not justified near the horizon although we need them to relate the global and AdS-Rindler modes. 
It is manifest if we consider the bulk wave packet localized on the geodesics (that we called the horizon-to-horizon geodesics in \cite{Sugishita:2022ldv}) leaving on the past horizon and passing through the future horizon in the AdS-Rindler patch $D(M_A)$ because the semi-classical description for the AdS-Rindler observer is not valid at the stretched horizon. 
Thus, we cannot trust the relation between the global and AdS-Rindler modes when the bulk free description is not UV complete.

\paragraph{Contradiction with the quantum corrected Ryu-Takayanagi formula}
We can also argue that the entanglement wedge reconstruction contradicts the FLM proposal \cite{Faulkner:2013ana} that the CFT entanglement entropy for the subregion $A$ agrees with the area term (Ryu-Takayanagi formula) including the back reaction and the bulk entanglement entropy for the $\mathcal{O}(N^0)$ order in the $1/N$ expansion.
If we consider an excited state $\ket{\Psi}$ where the expectation value of the CFT stress-energy tensor $\bra{\Psi}T_{\mu\nu}\ket{\Psi}=\mathcal{O}(N^0)$ and compare the difference of the entropy between $\ket{\Psi}$ and the vacuum state $\ket{0}$, the FLM proposal means that 
\begin{align}
\label{FLMeq}
    \delta S_A^\text{CFT}=\frac{\delta A_\Sigma}{4 G_N}+\delta S_{M_A}^\text{bulk}
\end{align}
holds for the leading order of the $1/N$ expansion, where $\delta S_A^\text{CFT}$ is the difference between the entanglement entropy for $\ket{\Psi}$ and $\ket{0}$ for subregion $A$ in CFT, $\delta S_{M_A}^\text{bulk}$ is that for bulk subregion $M_A$, and $\delta A_{\Sigma}$ is the change of the area of the Ryu-Takayanagi surface $\Sigma$ for the subregion $A$ due to the back reaction by the excitation.
Note that the entanglement entropy itself is UV divergent but the differences between the excited states and the vacuum are UV finite. 

As an excited state $\ket{\Psi}$, let us take the bulk coherent state $\ket{\Psi}=e^{B}\ket{0}$ for the free bulk scalar field corresponding to an arbitrary classical configuration $\phi_{cl}$ (see Appendix~\ref{App:coherent} for details).
The operator $e^{B}$ is unitary and also factorized in the bulk as $e^{B}=e^{B_{M_A}}e^{B_{M_{\bar{A}}}}$ where $B_{M_A}$ and $B_{M_{\bar{A}}}$ are supported only on the bulk subregion $M_A$ and $M_{\bar{A}}$ respectively. 
Then, the bulk entanglement entropy for $\ket{\Psi}$ is the same as that for the vacuum and  $\delta S_{M_A}^\text{bulk}=0$. 
The area term $\delta A_{\Sigma}$ is computed by solving the Einstein equation including the contribution of the classical bulk stress tensor $T_{\mu\nu}^\text{bulk}[\phi_{cl}]$ corresponding to the classical configuration $\phi_{cl}$ which is computed as $T_{MN}^\text{bulk}[\phi_{cl}]= \bra{\Psi}T_{MN}^\text{bulk}\ket{\Psi}-\bra{0}T_{MN}^\text{bulk}\ket{0}$. 
Since the classical configuration is arbitrary, we expect that there are cases where the back-reaction deforms the area of the Ryu-Takayanagi surface $\delta A_{\Sigma} \neq 0$.
Note that the correction $\delta A_{\Sigma}$ is $\mathcal{O}(G_N)$ and thus $\delta A_{\Sigma}/(4G_N)$ is $\mathcal{O}(G_N^0)$ or $\mathcal{O}(N^0)$.
Thus, the right-hand side of \eqref{FLMeq} can be non-vanishing values at the leading order $\mathcal{O}(N^0)$.

On the other hand, if the entanglement wedge construction holds, the bulk operators $B_{M_A}$ and $B_{M_{\bar{A}}}$ can be represented as CFT operators supported on $A$ and $\bar{A}$ respectively. 
Thus, the operators $e^{B_{M_A}}$ and $e^{B_{M_A}}$ are the unitary operators on $A$ and $\bar{A}$. 
It leads to $\delta S_A^\text{CFT}=0$ that contradicts \eqref{FLMeq}. 
Therefore, the entanglement wedge is incompatible with the FLM proposal. 
As we stressed in the previous subsection, we may encounter many contradictions when we consider subregions if we trust the $1/N$ expansion (or the generalized free theory with the higher-order $1/N$ perturbative corrections).\footnote{
The discussion here is completely different from the argument in \cite{Akers:2020pmf} that a naive “quantum-corrected” RT formula can fail. \cite{Akers:2020pmf} focuses on a special situation in which, besides the minimal RT surface, there exists another candidate surface whose area differs from the minimal one only by an $\mathcal{O}(N^0)$ amount, and considers a mixture of states. In our setup, by contrast, we take the boundary subregion in the CFT to be a single connected ball-shaped region and do not consider a mixture of states, and thus no such subtlety arises.

Working to first order in a perturbation where the difference between $\ket{\Psi}$ and $\ket{0}$ is $\mathcal{O}(N^0)$, evaluating the bulk-matter contribution and the shift of the RT surface due to backreaction separately is precisely what it means to implement the QES prescription at this order. Any additional correction to the bulk entropy coming from the fact that the classical RT surface itself is displaced appears only at higher orders in the $1/N$ expansion. Our point is that an inconsistency already shows up at the leading order ($\mathcal{O}(N^0)$) in the $1/N$ expansion.}

\paragraph{A simple bulk reconstruction picture}

As we will see below,
we can identify which part of the bulk local operator supported on $M_A$ cannot be reconstructed from the CFT operators supported on $A$
as done in \cite{Terashima:2020uqu, Terashima:2021klf, Terashima:2023mcr} using the results in \cite{Terashima:2017gmc} \cite{Terashima:2019wed}.
Here, the bulk local operator is defined by the global HKLL reconstruction.
In these papers, the bulk wave packet state was constructed from the (smeared) bulk local operators.
Because it always reaches the asymptotic boundary by the backward (global) time-evolution and the BDHM relation in the global patch,
it is represented by the time-evolution of the states given by acting smeared CFT primary operator with almost fixed energy and momentum on the vacuum at a point on the boundary, say $x$.
Although the wave packet is localized on a (light-like) curve in the bulk, the time evolution of the state in the CFT picture is similar to the two light-like particles (in the two-dimensional CFT)
and the VEV of the energy-momentum tensor is localized around the two points on a time slice.
Thus, if $x \in D(A)$, this state is supported in $A$ because the two points stay in $D(A)$.
However, if $x \notin D(A)$ the state is not supported in $A$, and the bulk wave packet is localized on the horizon-to-horizon geodesics. 
For the states corresponding to the wave packets localized on the horizon-to-horizon geodesics, \eqref{ewr} does not hold.\footnote{
More precisely, if we consider \eqref{ewu01} where
$\tilde{\phi}^R$ is the CFT operator for the bulk wave packet, the violation is clear because the VEV of the energy-momentum tensor at the point outside $D(A)$ should vanish.
}
Note that the dominant modes for the wave-packet localized on the horizon-to-horizon geodesics correspond to the ``tachyonic'' modes for the CFT on the Rindler wedge $D(A)$ which should be absent in the finite $N$ CFT \cite{Sugishita:2022ldv}.
The problem of the tachyonic modes is owing to the use of the semi-classical description near the horizon as argued above.

\section{AdS/CFT for subregion} 
\label{sec:comp}
In the AdS/CFT correspondence, it is expected that if we take (at least for a simple) subregion $A$ of CFT, the gravitational theory on a subregion $M_A$, whose boundary consists of $A$ and the Ryu-Takayanagi surface of $A$, in the bulk is dual to the CFT on $A$.
However, we have seen that the subregion duality and the causal (and also entanglement) wedge reconstruction are not valid, even at the leading order of the $1/N$ corrections. 
Thus the above expectation seems to be invalid.
Nevertheless, the AdS/CFT correspondence for the subregion may be correct in the following sense at least for the AdS-Rindler case:
The large $N$ leading order of any (global) vacuum correlation functions of the low-energy operators supported only in $M_A$ can be reproduced from the CFT correlation functions on $A$.\footnote{
Here, the operators supported only in $M_A$ mean the operators that are contained in the bulk theory of the AdS-Rindler patch. These may be 
the gauge invariant operators supported only in $M_A$
without gauge fixing.
}
Nevertheless, if we consider two different bulk subregions $M_A$ and $M_B$, the bulk operator $\phi(x)$ with $x \in M_A \cap M_B$ reconstructed by the HKLL from $A$ and $B$ are different.  
We will show it in two different ways: using the bulk picture and AdS/CFT correspondence for the Poincare patch.
Though the discussion in this section may be essentially known, it clarifies its relationship with what was discussed in the previous section.


Let us consider the vacuum state $\sigma = \ket{0}\bra{0}$ in the CFT on $S^{d-1}$ with the global Hamiltonian.
For the Rindler subregion $A$, the reduced density matrix is $\sigma_A=\tr_{\bar{A}} \sigma$.
In the bulk picture, the vacuum state $\sigma = \ket{0}\bra{0}$ is as the global AdS vacuum $\ket{0} \bra{0}$.
We can consider the corresponding bulk subregion $M_A$ and 
define the reduced density matrix on $M_A$ as $\sigma_{M_{A}}=\tr_{M_{\bar{A}}} \sigma$.
(This partial trace is not well-defined because the bulk spacetime description is only an approximation,
and then $\sigma_{M_{A}}$ makes sense only for the low energy approximation.
Furthermore, the bulk Hilbert space may not be factorized even approximately and such partial trace may not be well-defined for finite $N$.)

\begin{figure}[htbp]
\centering
\includegraphics[width=8cm]{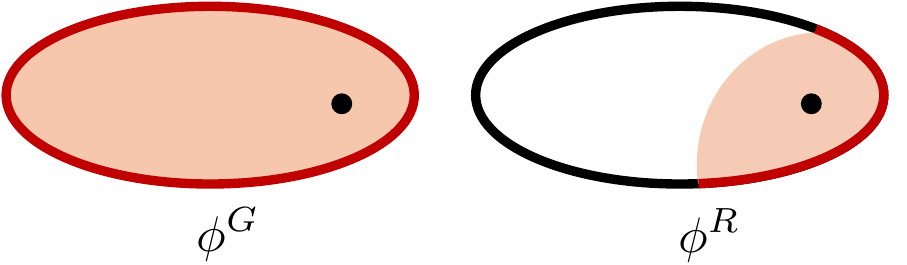}
\caption{Two CFT operators $\phi^G(X,\mathcal{O}^\text{CFT})$ and $\phi^R(X,\mathcal{O}^\text{CFT})$ have different support but reproduce the same bulk two-point functions. Thus, they have the same correlation functions in the large $N$ limit.
}
\label{fig:phiG-phiR}
\end{figure}

The above statement of the AdS/CFT correspondence for subregion is that for any (smeared) bulk local operator $\phi(X)$ on the subregion ($X\in D(M_A)$) there exists the CFT operator $\phi^R(X,\mathcal{O}^\text{CFT})$ supported only in CFT subregion $A$ such that we have
\begin{align}
    \bra{0} \phi(X_1) \dots \phi(X_n)\ket{0} =\tr_A (\sigma_A \phi^R(X_1,\mathcal{O}^\text{CFT}) \dots \phi^R(X_n,\mathcal{O}^\text{CFT}))
    \label{cfr1}
\end{align}
in the large $N$ limit ($N=\infty$) where all points $X_i$ are in $D(M_A)$ except for the stretched horizon region (neighborhood of the boundary of $D(M_A)$).
Note that $\phi^R(X,\mathcal{O}^\text{CFT})$ is different from the global HKLL reconstruction of the bulk operator $\phi(X)$ that we denote by $\phi^G(X,\mathcal{O}^\text{CFT})$ whose support is the entire space $S^{d-1}$ (see Fig.~\ref{fig:phiG-phiR}).
It should be emphasized that \eqref{cfr1} is valid at the leading order of large $N$. In other words, the equation is at the level of a (generalized) free theory, where higher-order correlation functions factorize into two-point functions.
As we stated in the previous subsection, the subregion duality does not hold and 
thus
\begin{align}
\label{phi-neq}
\phi^G(X,\mathcal{O}^\text{CFT}) \ket{\psi}\neq \phi^R(X,\mathcal{O}^\text{CFT})\ket{\psi},
\end{align}
when we include the leading $1/N$ corrections
where $\ket{\psi}$ are generic low-energy states, even if we include $1/N$ corrections to the reconstructed operators $\phi^G$ and $\phi^R$.
Both states in \eqref{phi-neq} are different in the entire space as can be confirmed by looking at the energy distribution, which can be represented as the higher point correlation function, as done in the previous section. 
This means that the ``local'' operator $\phi(X)$ 
given by the AdS-Rindler HKLL reconstruction is 
different from the one given by the global HKLL reconstruction as stated in the previous section. 
Thus, the two CFT operators $\phi^G(X_1,\mathcal{O}^\text{CFT})$ and $\phi^R(X_1,\mathcal{O}^\text{CFT})$ describe 
the same two-point functions
in the bulk subregion $M_A$ 
although they are different in the true finite $N$ theory.
This sounds similar to the black hole complementarity \cite{Susskind:1993if} in the sense that there are different but consistent descriptions, depending on the observers, for the event in the subregion (the outside of the ``horizon'').
We call the concept the subregion complementarity.

\subsection{Using bulk picture} 

First, we will consider the free limit of the bulk theory as the UV complete theory 
although only at the low energy the semi-classical theory is a good approximation of the finite $N$ holographic CFT.
Then, as explained above, the bulk local operator\footnote{More precisely we here consider the smeared operators so that they contain only low-energy modes, and the smearing functions are supported only in $M_A$.} $\phi^R(X)$ in the AdS-Rindler quantization in the AdS-Rindler subregion $D(M_A)$ 
is identical to the bulk local operator $\phi(X)$ in the global-time quantization at the same point in the subregion in the entire spacetime.
This implies that the correlation functions are the same, i.e.
\begin{align}
   \bra{0} \phi(X) \phi(X') \ket{0} 
   =\tr_{M_A} 
   (\sigma_{M_A} \phi^R(X) \phi^R(X') ) ,
\end{align}
where 
 $X, X' \in D(M_A)$.
Then, if we define $\mathcal{O}(x) $ as the boundary limit of $\phi(X) (= \phi^R(X))$ with an appropriate scaling factor, 
$\mathcal{O} (x) $ gives the correct two-point function of the corresponding scalar of CFT.
Here, $\phi^R(X)$ is written by the AdS-Rindler creation/annihilation modes and 
$\mathcal{O}(x) $ can be written by them also.
$\phi(X)$ is written by the global AdS creation/annihilation modes and 
$\mathcal{O} (x) $ can be written by them also.
Using the global AdS modes, we can rewrite $\phi(X)$ as a functional of $\mathcal{O} (x) $,
which is given, for example, by the global HKLL bulk reconstruction formula,
and we will denote it as $\phi^G(X, \mathcal{O})$.
On the other hand, using the AdS-Rindler modes, 
we can rewrite $\phi^R(X)$ as a functional of $\mathcal{O} (x) $,
which is given, for example, by the AdS-Rindler HKLL bulk reconstruction formula,
and we will denote it as $\phi^R(X, \mathcal{O})$.
Note that $\phi^R(X, \mathcal{O})$ is  linear in $\mathcal{O}(x)$ and the support is in $A$. Thus, we have
\begin{align}
   \tr_{M_A} 
   (\sigma_{M_A} \phi^R(X) \phi^R(X') ) 
   =\bra{0}
    \phi^R(X,\mathcal{O}) \phi^R(X',\mathcal{O}) \ket{0}=\tr_{A} 
   (\sigma_{A} \phi^R(X,\mathcal{O}) \phi^R(X',\mathcal{O}) ).
   \label{cfr}
\end{align}
Because the vacuum two-point function is determined only by the conformal dimension,
we should have
\begin{align}
\tr_{A} 
   (\sigma_{A} \phi^R(X,\mathcal{O}) \phi^R(X',\mathcal{O}) )=
\tr_{A} 
   (\sigma_{A} \phi^R(X,\mathcal{O}^\text{CFT}) \phi^R(X',\mathcal{O}^\text{CFT}) )
\end{align}
where $\mathcal{O}^\text{CFT} (x) $ is the primary scalar operator of the finite $N$ CFT.\footnote{
The AdS-Rindler modes contain ``tachyonic'' creation/annihilation modes.
For $\phi^R(X,\mathcal{O}^\text{CFT})$, these correspond to the Fourier transformation of  $\mathcal{O}^\text{CFT}(x)$ 
with the tachyonic energy momenta in the Rindler coordinates on $A$.
Such Fourier modes $\mathcal{O}^\text{CFT}(x)$ vanish around the Rindler vacuum, but 
do not vanish around the finite temperature state.
}
Thus, for the two-point function at $X, X' \in D(M_A)$, we have
\begin{align}
   \bra{0} \phi(X) \phi(X') \ket{0} 
   =\tr_{A} 
   (\sigma_{A} \phi^R(X,\mathcal{O}^\text{CFT}) \phi^R(X',\mathcal{O}^\text{CFT}) ).
   \label{cf1}
\end{align}
This also means that  
the bulk correlation functions can be reproduced both from the global and the Rindler reconstructions:
\begin{align}
   \bra{0} \phi(X) \phi(X') \ket{0} 
   =\bra{0} \phi^G(X,\mathcal{O}^\text{CFT}) \phi^G(X',\mathcal{O}^\text{CFT})\ket{0}
      = \bra{0} \phi^R(X,\mathcal{O}^\text{CFT}) \phi^R(X',\mathcal{O}^\text{CFT})\ket{0}.
   \label{cf2}
\end{align}
Multi-point correlation functions can also be reconstructed in the large $N$ and low-energy limit because at the leading order the bulk theory is free and the large $N$ factorization holds in the CFT side.
Therefore, in the sense that we can reproduce the bulk correlation functions on the global vacuum in this limit, the AdS-Rindler HKLL bulk reconstruction formula is correct and the AdS/CFT correspondence for the subregion holds.

It is important to note that, even though \eqref{cf1} is correct, 
$\phi^G$ and $\phi^R$ are different CFT operators for finite $N$ as 
\begin{align}
   \phi^G(X,\mathcal{O}^\text{CFT})  \neq \phi^R(X,\mathcal{O}^\text{CFT}) ,
   \label{ew0}
\end{align}
even in the low energy approximation as we remarked above.
More precisely, for the smeared bulk local operators defined as
\begin{align}
   \tilde{\phi}^G(\mathcal{O}^\text{CFT}) =\int dX' K(X') \phi^G(X',\mathcal{O}^\text{CFT}), \,\,\,  
    \tilde{\phi}^R (\mathcal{O}^\text{CFT})=\int dX' K(X') \phi^R(X',\mathcal{O}^\text{CFT}),
   \label{ew01}
\end{align}
where $K(X')$ is a smearing function (or distribution) supported on $M_A$, 
\begin{align}
   \tilde{\phi}^G(\mathcal{O}^\text{CFT}) \ket{\psi} =  \tilde{\phi}^R (\mathcal{O}^\text{CFT}) \ket{\psi},
   \label{ew1}
\end{align}
does not hold for generic low-energy states $\ket{\psi}$ as we saw in the previous section.
We remark that the width of the smearing function $K(X')$ must be much greater than the cut-off length (i.e., the Planck length) and thus the center of $K(X')$ is further away from the AdS-Rindler horizon than the Planck length, and \eqref{cf2} is not valid near the horizon.

The invalidity of \eqref{ew1} is important because according to the entanglement wedge reconstruction (more precisely causal wedge reconstruction which is a special case of entanglement wedge reconstruction),
$\phi^R(X,\mathcal{O}^\text{CFT})$ which satisfies \eqref{ew1} should exist.
Instead, in the low energy approximation we have 
\begin{align}
   \bra{0} \phi^G(X,\mathcal{O}^\text{CFT}) \phi^G(X',\mathcal{O}^\text{CFT}) \ket{0} 
   = \tr_{A} 
   (\sigma_{A} \phi^R(X,\mathcal{O}^\text{CFT}) \phi^R(X',\mathcal{O}^\text{CFT}) ) .
   \label{ew2}
\end{align}
The subregion duality will also lead to 
\eqref{cf1} and \eqref{ew2}.
Here, we claim that \eqref{cf1} and \eqref{ew2} can be realized
although the causal wedge reconstruction (and thus the entanglement wedge reconstruction) is not correct in the holographic CFT.
It should be emphasized once again that \eqref{ew2} is restricted to two-point functions; for three-point (and higher-order) functions, such an expression no longer holds. This, of course, is consistent with the assertion made in \eqref{ew0}.


This difference between $\phi^G(X,\mathcal{O}^\text{CFT})$ and $ \phi^R(X,\mathcal{O}^\text{CFT})$ is the observer dependence (the patch and foliation dependence). 
It seems similar to the state dependence of the bulk reconstruction \cite{Papadodimas:2013jku} but not the same because we consider only the global vacuum state.
Note that the description in the AdS-Rindler patch using $ \phi^R(X,\mathcal{O}^\text{CFT}) $ is 
not valid near the Rindler horizon because of the finite $N$ effect as we have discussed
although the description in the global coordinate using $ \phi^G (X,\mathcal{O}^\text{CFT})$ is valid, of course.

We call the existence of the two different descriptions depending on the observers, the subregion complementarity. 
The CFT on a subregion has a gravity dual on the corresponding subregion which is 
consistent including $1/N$ corrections
even though the bulk ``local'' operators in the theory on the subregion differ from the theory on the entire space (or another subregion) except for the $N=\infty$ case.
It might be similar to the black hole complementarity \cite{Susskind:1993if}.
$\phi^G$ corresponds to an observer passing through the horizon and $\phi^R$ does to an external observer.

It is interesting to extend this discussion to more general subregions. 
We have considered a single spherical subregion (whose size is arbitrary) in the sphere.
It is interesting to take $A$ as the union of several spherical subregions.
In this case, the bulk correlation functions on the union of the bulk causal wedges of the boundary subregions can be reproduced by the AdS-Rindler type bulk reconstructions for each subregion.
The entanglement wedge will play an important role in this analysis although we do not have a concrete one for now.
It is important to confirm the subregion complementarity holds for general causal (or entanglement) wedges in the AdS/CFT correspondence as the quantum gravity.

\subsection{Using AdS/CFT for the Poincare and Hyperbolic patches} 
\label{sec:Poincare}

Here, we will provide another derivation of the above statement of the AdS/CFT correspondence for the subregion, i.e. bulk correlators in $D(M_A)$ can be reconstructed from the CFT on subregion $D(A)$,  using the Poincare and hyperbolic patches of AdS. 

The cylinder metric on subregion $D(A)$ is conformal to that for $\mathbf{R} \times \mathbf{H}^{d-1}$ as
\begin{align}
    -d\tau^2+d\Omega_{d-1}^2=e^{2\Phi}\left(-dt_R^2+ dH_{d-1}^2\right),
\end{align}
where $\tau$ is the global time for cylinder and $t_R$ is the Rindler. 
The conformal factor $e^{2\Phi}$ is given by \eqref{confac-gl-Rind} in Appendix~\ref{app:AdSRindler}.
Thus, CFT on $D(A)$ is the same as that on $\mathbf{R} \times \mathbf{H}^{d-1}$.

We now focus on $d=2$. 
In particular, for $d=2$, $\mathbf{R} \times \mathbf{H}^{d-1}$ is nothing but the Minkowski space $\mathbf{R}^{1,1}$ with time $t_R$. 
If the state is the vacuum for the Rindler Hamiltonian, the dual bulk geometry is the three-dimensional Poincare AdS, and CFT on $D(A)$ is dual to gravity on the Poincare patch. 

Now we consider the vacuum state for the global Hamiltonian for time $\tau$.
On the subregion $A$, we consider the reduced density matrix $\rho_A$, which is a mixed state instead of the Rindler vacuum. 
More precisely, it is known to be the finite temperature state with $T=1/(2 \pi)$ for the Hamiltonian for time $t_R$ as the usual Unruh effect for this specific choice of the subregion $A$.
Thus, CFT on the subregion $A$ around the global vacuum is the same as CFT on the Minkowski space $\mathbf{R}^{1,1}$ around the finite temperature state.

Forgetting the entire space, we are just considering the holographic CFT on $\mathbf{R}^{1,1}$ at temperature $T=1/(2 \pi)$. 
The bulk dual should be the Poincare black hole geometry with temperature $T=1/(2 \pi)$ on the Poincare patch. 
Except for the stretched horizon region where the low-energy approximation breaks down, the (smeared) bulk correlation functions at $T=1/(2 \pi)$ on the outside of the horizon can be reproduced from the CFT on $D(A)$.
Furthermore, the black hole solution is known to be equivalent to the AdS-Rindler patch. 
Thus, the CFT on $D(A)$ around the mixed state $\rho_A$ describes the bulk theory on the AdS-Rindler patch.

The HKLL bulk reconstruction for the black hole solution is the AdS-Rindler bulk reconstruction.
Note that this discussion does not imply the bulk local operators reconstructed by the global and Rindler are the same.
They are completely different as mentioned in the previous subsection.


The discussion is similar for higher dimensions $d>2$.
If we take the state as the global vacuum, the reduced identity matrix on the spherical subregion $A$ is the finite temperature state with temperature $1/(2\pi)$ for the Hamiltonian for time $t_R$ \cite{Casini:2011kv}. 
Thus, the theory on $D(A)$ is the holographic CFT on $\mathbf{R} \times \mathbf{H}^{d-1}$ with temperature $1/(2\pi)$. 
The dual geometry should be the $(d+1)$-dimensional hyperbolic black hole. 
However, the hyperbolic black hole with temperature $1/(2\pi)$ is the same as the AdS-Rindler geometry (see, e.g., \cite{Emparan:1999gf} and also Appendix~\ref{app:AdSRindler}).
Thus the bulk theory on the AdS-Rindler patch can be described by the CFT on $\mathbf{R} \times \mathbf{H}^{d-1}$ around the finite temperature state at $T=1/(2\pi)$, that is, the CFT on $D(A)$ around the mixed state $\rho_A$.

\section{Black holes} 
\label{sec:bh}
\subsection{Eternal Black holes} 

It is known that the Rindler space is similar to the black hole by focusing on the near horizon region.
The observer with the Rindler time corresponds to the one in the region outside the black hole horizons, i.e. the static observer.
In particular, the eternal AdS black hole is similar to the empty AdS case.\footnote{Indeed, the AdS-Rindler space is a special case of the hyperbolic black holes in AdS as reviewed in Appendix~\ref{app:AdSRindler}, and the patch covers the outside region of the horizon.}
Indeed, the gravitational theory around the eternal black hole in the asymptotic AdS space-time is believed to be described by the thermofield double state in 
the tensor product of the identical two CFTs \cite{Maldacena:2001kr}.
If we trace out the degrees of freedom in either one of the CFTs, the state becomes the thermal state which corresponds to the region outside the horizon of the eternal black hole. 
This story is similar to the relation between the global AdS and the AdS-Rindler, where the Schwarzschild-type coordinates outside the horizon correspond to the AdS-Rindler coordinates.

In the bulk theory around the eternal AdS black hole, we can take a different coordinate system, for example, the Kruskal-type coordinates, in which the whole system is described, as we can take the global coordinates instead of the AdS-Rindler ones in the empty AdS space, although it is not clear that there exists the Hamiltonian corresponding to the Kruskal time in the tensor product of the finite $N$ CFTs.

We expect that the subregion complementarity also holds in the eternal black holes as in the empty AdS case.  
There are different CFT descriptions of the bulk spacetime outside the horizon depending on the observers (the Schwarzschild-type or the Kruskal-type).
For the static observer, which corresponds to either one CFT in the two copies, the region near the horizon cannot be described because the semi-classical approximation breaks down due to the finite $N$ effect as in the AdS-Rinlder case. 
However, the inside of the horizon can be described by the tensor product of the two CFTs, as in the entire CFT on $S^{d-1}$ for the vacuum case.

\subsection{Single-sided black holes} 

The situation for single-sided black holes is drastically different from the eternal black holes.

Let us consider the black hole in the AdS spacetime that comes from the gravitational collapse, sometimes called the single-sided black hole compared with the eternal black hole.
In this case, the dual theory is the single CFT, and the initial state is a pure state corresponding to the bulk object before its collapse. 
It evolves in time to a typical pure state with the temperature fixed by the energy of the initial state, which corresponds to the black hole in the bulk.\footnote{
Here, we assume that the black hole is the large AdS-Black hole, which can be in equilibrium with the thermal gas around it by the Hawking radiation.
For the large AdS-Black hole, the discussion here can be applied except that it is not a thermal state.
}
This is in the description for the static observer outside the horizon.
The semi-classical description is invalid near the horizon as for the AdS-Rindler observer in the empty AdS.
This is consistent with the CFT picture, in which the thermal state is in the deconfinement phase of the gauge theory, and there are many ($\mathcal{O} (N^2) $) light modes.
These come from the brick wall of the black hole \cite{tHooft:1984kcu} in AdS/CFT \cite{Iizuka:2013kma}.
In contrast to the eternal black hole, in this case, 
the Hilbert space is for the single CFT and then
it may be impossible to have another description where the infalling observer will see the smooth spacetime near the horizon.

The absence of an infalling observer implies that there is no semi-classical description at the stretched horizon for single-sided black holes. 
In this respect, the equivalence principle might be violated for single-sided black holes in quantum gravity (= finite $N$ CFT). 
This is completely different from eternal black holes where we have semi-classical descriptions corresponding to infalling observers.
In eternal black holes, the static observer outside the horizon sees a mixed state. 
It can be purified by extending the Hilbert space.
This extended description corresponds to the infalling observer. 
On the other hand, for the single-sided case, the state is already pure,\footnote{We consider a finite $N$ CFT on the compact space $S^{d-1}$. The operator algebra should be type I and there exist pure states.} with no room for extending the Hilbert space.\footnote{If we further perform coarse-graining for the pure state, we may obtain a mixed state and can consider the purification (for such attempts, see for example \cite{Engelhardt:2018kcs, Chandra:2022fwi, Kudler-Flam:2025cki, Mori:2025jej}). Our discussion here does not suppose such a coarse-graining (or consider only coarse-grainings that preserve the purity of our low-energy pure state).} Even if we could extend it, any added degrees of freedom would not entangle with the original ones and thus be completely decoupled. 
Thus, the black hole complementarity does not hold.  
As its name suggests, the subregion complementarity holds only when a subregion is under consideration.

The possibility of the violation of the equivalence principle has been proposed by Mathur as the fuzzball conjecture \cite{Mathur:2005zp, Mathur:2009hf} (and later it was also argued as the fire wall \cite{Almheiri:2012rt}).\footnote{
In \cite{Czech:2012be}, it was discussed that the fuzzball-like structure should appear in the typical state corresponding to the black hole.
}
We here expect a brick wall-like behavior near the stretched horizon, where the bulk spacetime picture is not valid 
because of the shortage of degrees of freedom and the bulk semi-classical theory cannot be applied. 
Note that this discussion can be applied to 
a (hypothetical) static heavy star equilibrium with the radiation around it such that the would-be horizon inside the star is very close to the surface of the star.
If the distance between them is the order of the Planck length, the shortage of degrees of freedom to reconstruct the bulk theory, like the brick wall model, occurs for the finite $N$ CFT \cite{Terashima:2020uqu}
as the black hole case. 
In this case, the inside of the star cannot be described by 
the bulk theory.

\section*{Acknowledgement}

SS thanks Sumit Das, Samuel Leutheusser, Gautam Mandal and Masamichi Miyaji for the helpful discussions.
SS acknowledges support from JSPS KAKENHI Grant Numbers JP 21K13927, 22H05115.
This work was supported by JSPS KAKENHI Grant Number 17K05414.
This work was supported by MEXT-JSPS Grant-in-Aid for Transformative Research Areas (A) ``Extreme Universe'', No. 21H05184.

\hspace{1cm}

\appendix

\section{General AdS-Rindler and hyperbolic black holes}
\label{app:AdSRindler}
\subsection{AdS-Rindler}
$(d+1)$-dim AdS space is embedded as 
\begin{align}
    -(X^{-1})^2 -(X^{0})^2 +(X^1)^2+\cdots +(X^d)^2 =-1.
\end{align}
The global coordinates $(\tau,\rho,\Omega)$ are obtained by parameterizing the embedding coordinates as 
\begin{align}
    X^{-1}=\frac{1}{\cos \rho}\cos \tau,\quad   X^{0}=\frac{1}{\cos \rho}\sin \tau,\quad
    X^i=\tan \rho\, \hat{x}^i(\Omega), 
\end{align}
The metric is 
\begin{align}
    ds^2=\frac{1}{\cos^2\! \rho}\left(-d\tau^2+d \rho^2+ \sin^2 \!\rho\, d\Omega_{d-1}^2\right).
\end{align}
We take the  spherical coordinates $\hat{x}^i(\Omega)$ as 
\begin{align}
    \hat{x}^1(\Omega)=\cos \theta,\quad
    \hat{x}^j(\Omega)=\sin\theta\,\hat{y}^j(\bO)\quad (j=2, \dots, d) \quad \text{with} \quad 0\leq \theta \leq \pi,
\end{align}
where $\hat{y}^j(\Omega)$ $(j=2,\dots,d)$ are the embedding of the sphere into $\mathbf{R}^{d-1}$.\footnote{For $d=2$, the range of $\theta$ is $-\pi \leq \theta \leq \pi$.}
The asymptotic boundary of the global coordinates is a cylinder $\mathbf{R}\times S^{d-1}$.

\begin{figure}[htbp]
\centering
\includegraphics[width=5cm]{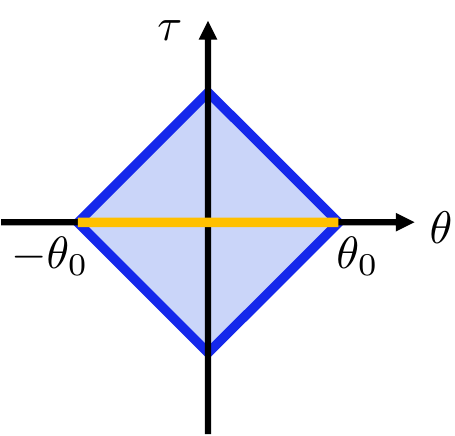}
\caption{Causal diamond $D(A)$ in $(\tau,\theta)$-coordinates in $d=2$. The yellow line represents subregion $A$, and the blue diamond region is $D(A)$.
}
\label{figapp:D(A)}
\end{figure}

The general AdS-Rindler coordinates are given by the following parametrization:
\begin{align}
\begin{split}
    &X^{-1}=\frac{\sqrt{1+\xi^2}\cosh \chi+ \cos \theta_0 \xi \cosh t_R}{\sin \theta_0}, \quad X^0=\xi \sinh t_R,\\
    &X^1=\frac{ \xi \cosh t_R+ \cos \theta_0\sqrt{1+\xi^2}\cosh \chi}{\sin \theta_0},  \quad
    X^j=\sqrt{1+\xi^2} \sinh \chi\, \hat{y}^j(\bO)\quad (j=2, \dots, d),
\end{split}
\end{align}
where $\theta_0$ is a constant in the range $0<\theta_0 \leq \pi$.
The range of $\chi$ is $0\leq \chi <\infty$ for $d\geq 3$ and $-\infty < \chi <\infty$ for $d=2$. 
The coordinates cover the bulk causal wedge $D(M_A)$ associated with the spherical cap ($0\leq |\theta|\leq \theta_0$) subregion $A$ on the boundary. ($t_R=0,\,  \xi=0$ surface is the Ryu-Takayanagi surface anchored to the $\theta=\theta_0$ entangling surface at $\tau=0$ on the boundary.)
If $\theta_0 = \pi/2$, the $t_R=0$ surface covers the half region on the $\tau=0$ surface ($\tau$ is global time).
The asymptotic boundary of $D(M_A)$ is the causal diamond $D(A)$, which is the spacetime subregion in the cylinder $\mathbf{R}\times S^{d-1}$. 
For example, in $d=2$, $D(A)$ is given as Fig.~\ref{figapp:D(A)} (see also Fig.~\ref{fig:D(A)}).

In these coordinates, the metric becomes
\begin{align}
\label{AdSRind-metric}
    ds^2=-\xi^2 dt_R^2+\frac{d\xi^2}{1+\xi^2}+(1+\xi^2) dH_{d-1}^2,
\end{align}
where $dH_{d-1}^2= d\chi^2 +\sinh^2\chi d\Omega_{d-2}^2$ is the metric of $(d-1)$-dimensional hyperbolic space $\mathbf{H}^{d-1}$ (for $d=2$, $\mathbf{H}^{1}=\mathbf{R}$ in our convention.).
The asymptotic boundary of the AdS-Rindler coordinates is $\mathbf{R}\times \mathbf{H}^{d-1}$.

On the general AdS-Rindler wedge, we can express the global coordinates $(\tau,\rho,\theta,\bO)$ in terms of the AdS-Rindler coordinates  $(t_R,\xi,\chi, \bO)$ as
\begin{align}
\begin{split}
\tan \tau &=\frac{\sin \theta_0 \xi\sinh t_R}{\sqrt{1+\xi^2}\cosh \chi+\xi\cos\theta_0 \cosh t_R},
\\
    \cos\rho &=\frac{1}{\sqrt{\left(\frac{\sqrt{1+\xi^2}\cosh \chi+ \cos \theta_0 \xi \cosh t_R}{\sin \theta_0}\right)^2+\xi^2 \sinh^2 t_R}},
    \\
    \tan \theta &=\frac{\sin \theta_0 \sqrt{1+\xi^2}\sinh \chi}{\xi\cosh t_R+ \cos \theta_0\sqrt{1+\xi^2}\cosh \chi}.
\end{split}
\end{align}

On the asymptotic boundary of general AdS-Rindler, $(\tau, \theta)$ are related with $(t_R, \chi)$ as
\begin{align}
\label{bd-glob-Rind}
\tan \tau =\frac{\sin\theta_0\sinh t_R}{\cosh \chi+\cos\theta_0 \cosh t_R}, \quad
    \tan \theta =\frac{\sin\theta_0\sinh \chi}{\cosh t_R+\cos\theta_0 \cosh \chi}.
\end{align}
This coordinate transformation is the same as given in \cite{Casini:2011kv}.
The cylinder metric $\mathbf{R}\times S^{d-1}$ is conformally mapped to that for $\mathbf{R}\times \mathbf{H}^{d-1}$ as 
\begin{align}
    -d\tau^2+d\Omega_{d-1}^2=e^{2\Phi}\left(-dt_R^2+ dH_{d-1}^2\right),
\end{align}
where the conformal factor $e^{2\Phi}$ is given by
\begin{align}
\label{confac-gl-Rind}
    e^{2\Phi}=\frac{\sin^2 \theta_0}{(\cosh\chi +\cos\theta_0 \cosh t_R)^2+\sin^2 \theta_0 \sinh^2 t_R}.
\end{align}
For example, for $d=2$, the subregion $D(A)$ in $(\tau, \theta)$-plane in Fig.~\ref{figapp:D(A)} is conformally mapped to $\mathbf{R}^{1,1}$ parametrized by $(t_R,\chi)$ by the map \eqref{bd-glob-Rind}.

When $\theta \sim \pi$, that is, the AdS-Rindler wedge covers almost the entire space covered by the Poincare patch, it is more convenient to rescale the coordinates as $t_R= (\sin \theta_0) t$, $\chi= (\sin \theta_0) r$. Then, taking the entire limit $\theta \to \pi$ with only keeping the finite $t$ and $r$, the relations \eqref{bd-glob-Rind} become 
\begin{align}
\label{bd-glob-Poin}
\tan \tau =\frac{2 t}{1-t^2+r^2}, \quad
    \tan \theta =\frac{2 x}{1+t^2-r^2}.
\end{align}
This is actually the conformal map between 
$\mathbf{R}\times S^{d-1}$ and $\mathbf{R}\times \mathbf{R}^{d-1}$.\footnote{The inverse is 
\begin{align}
    t=\frac{\sin \tau}{\cos \tau + \cos \theta }, \,\,\,
    r=\frac{\sin \theta }{\cos \tau + \cos \theta }.
\end{align}}
In this limit we have $-dt_R^2+ dH_{d-1}^2\sim \sin^2\theta_0(-dt^2+ dr^2+r^2 d\Omega_{d-2}^2)$, and the conformal factor \eqref{confac-gl-Rind} becomes
$e^{2\Phi}\sim \frac{1}{\sin^2\theta_0}\frac{4}{(1+(t-r)^2)(1+(t+r)^2)}$. $\frac{4}{(1+(t-r)^2)(1+(t+r)^2)}$ is exactly the conformal factor for the map \eqref{bd-glob-Poin}.
Note also that the AdS-Rindler metric \eqref{AdSRind-metric} becomes the Poincare metric
\begin{align}
    \frac{1}{z^2}(-dt^2+dz^2+dr^2+r^2 d\Omega_{d-2}^2),
\end{align}
by taking the limit $\theta_0\to \pi$ with the rescaling $t_R= (\sin \theta_0) t$, $\xi=\frac{1}{\sin\theta_0 z}$, $\chi= (\sin \theta_0) r$ and taking only finite $t, z, r$ region.

\subsection{General hyperbolic black holes}
The geometry with metric \eqref{AdSRind-metric} represents a hyperbolic black hole with temperature $1/(2\pi)$. 
The AdS hyperbolic black holes with arbitrary temperature are given by (see, e.g., \cite{Emparan:1999gf})
\begin{align}
\label{hyper-BH-metric}
    ds^2 =-f(r) d t^2 +\frac{dr^2}{f(r)}+r^2 dH_{d-1}^2,
\end{align}
where
\begin{align}
    f(r)=r^2-1-\frac{\omega^{d-2}}{r^{d-2}}.
\end{align}
Here, $\omega$ is a parameter related to the mass density\footnote{The ADM mass $M$ is proportional to the (infinite) volume $V$ of the hyperbolic space $\mathbf{H}^{d-1}$. The mass density $\varrho$ means $M/V$.} $\varrho$ of the black hole as
\begin{align}
    \varrho=\frac{d-1}{16\pi G_N}\omega^{d-2}.
\end{align}
The position of the horizon $r_h$ is determined by $f(r_h)=0$, i.e.,
\begin{align}
    r_h^{d-2}(r_h^2-1)=\omega^{d-2}.
\end{align}
The temperature of the black hole is given by
\begin{align}
    T=\frac{1}{4\pi}\left.\frac{\partial f}{\partial r}\right|_{r=r_h}
    =\frac{1}{4\pi r_h}(d r_h^2 -d+2).
\end{align}

The vanishing mass\footnote{The vanishing mass solution is not the lowest mass state. The lowest mass state is given by $\omega^{d-2}=-2 d^{-\frac{d}{2}}(d-2)^\frac{d-2}{2}$, which means $r_h=\sqrt{\frac{d-2}{d}}$, $\rho=-\frac{(d-1)d^{-\frac{d}{2}}(d-2)^\frac{d-2}{2}}{8\pi G_N}$ and $T=0$.} case ($\omega=0$) corresponds to the AdS-Rindler metric \eqref{AdSRind-metric} with the coordinate transformation $r^2=1+\xi^2$. It indeed has temperature $T=1/(2\pi)$.

The $d$-dimensional holographic CFT on $\mathbf{H}^{d-1}$ at finite temperature should correspond to the gravitational theory around the background \eqref{hyper-BH-metric}.
\section{Explicit mirror map for free fields}\label{App:mirror}
Here we will explicitly construct the mirror map in a continuum field theory and see the issues. 
We consider massless free scalar on two-dimensional Minkowski space $(T, X)$. 
We divide the $T=0$ time-slice ($-\infty < X<\infty$) into $X<0$ region and $X>0$ region, and call their domains of dependence the left and right Rindler wedge, respectively. 
In each Rindler wedge, we can take the following coordinates:
\begin{align}
    &T-X=-e^{-t_R+x_R},\quad  T+X=e^{t_R+x_R} \quad \text{in the right Rindler}.\\
    &T-X=e^{t_L-x_L},\quad  T+X=-e^{-t_L-x_L} \quad \text{in the left Rindler}.
\end{align}
The coordinates $(t_R, x_R)$ and $(t_L, x_L)$ are the Rindler coordinates, and we will sometimes omit the subscripts $R, L$ for simplicity.

In the right wedge, the metric is given by
\begin{align}
    -dT^2+dX^2=e^{2 x_R}(-dt_R^2+dx_R^2),
\end{align}
and the left wedge, it is 
\begin{align}
    -dT^2+dX^2=e^{-2 x_L}(-dt_L^2+dx_L^2).
\end{align}

Each metric is conformally flat in the Rindler coordinates, and thus the massless free scalar can be quantized as in Minkowski coordinates. 
Since we are considering free massless scalar, right-moving and left-moving modes are decoupled, and we here focus on the right-moving field.
In the right Rindler wedge, the right-moving field is expanded by the Rindler modes as 
\begin{align}
    \phi(x)=\int^{\infty}_0\frac{dk}{4\pi k}\left(a_k^R e^{-ik(t_R-x_R)}+a_k^{R\dagger} e^{ik(t_R-x_R)}\right).
\end{align}
The discussion is the same for the left-moving field. 
$a_k^{R}, a^{R\dagger}_k$ are the annihilation and creation operators for the right Rindler modes and they satisfy
\begin{align}
    [a_k^{R}, a^{R\dagger}_{k'}]=4\pi k\delta(k-k').
\end{align}
In the left Rindler wedge, we also have the operators for the left Rindler (right-moving) modes, $a_k^{L}, a^{L\dagger}_k$, and they also satisfy 
\begin{align}
    [a_k^{L}, a^{L\dagger}_{k'}]=4\pi k\delta(k-k').
\end{align}

Here we consider the thermofield double state for these left and right modes\footnote{We need an IR regulator in order to make the spectrum discrete. Here we consider a problem in UV and thus we do not touch further this point.}
\begin{align}
    \ket{TFD}=\mathcal{N}\sum_{n}e^{-\frac{\beta E_n}{2}}\ket{E_n}_L \otimes \ket{E_n}_R,
\end{align}
where $\mathcal{N}$ is a normalization factor, and $\ket{E_n}_{L/R}$ are Fock states of the Rindler modes with energy $E_n$ for the Rindler Hamiltonian. 
Note that $\beta$ should be $2\pi$ for the state $\ket{TFD}$ to be the ground state of the Hamiltonian for the Minkowski time $T$.

We will construct a mirror map $\mathsf{m}$ from operators in $\mathcal{A}_L\otimes 1_R$ to ones in $1_L\otimes \mathcal{A}_R$ such that $\mathcal{O}_L \in \mathcal{A}_L\otimes 1_R$ and $\mir(\mathcal{O}_L) \in 1_L\otimes \mathcal{A}_R$ act on $\ket{TFD}$ as if they are the same operator, where $\mathcal{A}_{L, R}$ denotes the set of operators on the left or right Fock space, respectively. 
We can confirm that acting $a_k^{L}$ (more precisely $a_k^{L}\otimes 1_R$) to the state $\ket{TFD}$ is the same as acting $e^{-\frac{k}{2}\beta}a_k^{R\dagger}$ as
\begin{align}
    a_k^{L}\ket{TFD}=e^{-\frac{k}{2}\beta}a_k^{R\dagger}\ket{TFD}.
\end{align}
We similarly have
\begin{align}
    a_k^{L\dagger}\ket{TFD}=e^{\frac{k}{2}\beta}a_k^{R}\ket{TFD}.
\end{align}
Therefore, we have
\begin{align}
\label{mir-b}
    \mir( a_k^{L})=e^{-\frac{k}{2}\beta}a_k^{R\dagger}, \quad \mir(a_k^{L\dagger})=e^{\frac{k}{2}\beta}a_k^{R}.
\end{align}
If we define $a_{-k}^{L,R}=a_{k}^{L,R \dagger}$, the two equations in \eqref{mir-b} is unified as $\mir(a^{L}_{k})=e^{-\frac{k}{2}\beta}a_{-k}^{R}$.
Note that the map $\mir$ depends on the choice of the reference state $\ket{TFD}$ as $\beta$ explicitly appears in \eqref{mir-b}.
The mirror of multiple products of ladder operators is also obtained as
\begin{align}
\label{multi-mir}
    \mir(a_{k_1}^L \cdots a_{k_n}^L)=e^{-\frac{\beta}{2}\sum_i^n k_i}a_{-k_n}^R\cdots a_{-k_1}^R
\end{align}
as can be understood from the following equations 
\begin{align}
    a_{k_1}^L \cdots a_{k_n}^L\ket{TFD}=a_{k_1}^L \cdots a_{k_{n-1}}^L e^{-\frac{k_n}{2}\beta}a_{-k_n}^{R}\ket{TFD}=e^{-\frac{k_n}{2}\beta}a_{-k_n}^{R}a_{k_1}^L \cdots a_{k_{n-1}}^L \ket{TFD},
\end{align}
where we have used $a_{k}^{R}$ commutes with $a_k^L$.

If the local operator on the left Rindler wedge acts on $\ket{TFD}$, the excited state can also be created \textit{formally} by acting an operator supported on the Right wedge as
\begin{align}
    \phi_L(t,x) \ket{TFD}&= \int^{\infty}_0\frac{dk}{4\pi k}\left(a_k^L e^{-ik(t-x)}+a_k^{L\dagger} e^{ik(t-x)}\right)\ket{TFD}
    \nn
    &=
    \int^{\infty}_0\frac{dk}{4\pi k}\left(a_k^{R\dagger} e^{-ik(t-i\frac{\beta}{2}-x)}+a_k^{R} e^{ik(t-i\frac{\beta}{2}-x)}\right)\ket{TFD}
    \nn
    &=
    \phi_R(-t+i\beta/2,-x)\ket{TFD}.
    \label{local-mirror}
\end{align}
Here we put subscripts $L, R$ on $\phi$ to make it obvious that they are operators supported on the left or right wedge.
Thus, the mirror operator of $\mir(\phi_L(t,x))=\phi_R(-t+i\beta/2,-x)$.
We remark that this is just a formal relation because $\phi_R(-t+i\beta/2,-x)$ is not well-defined due to the blowing up factor $e^{k \beta/2}$ in the momentum representation. We also note that the above mirror map is different from the Tomita-Takesaki modular conjugation.

The position $(t_R,x_R)=(-t+i\beta/2,-x)$ with $\beta=2\pi$ is $(T-X, T+X)=(e^{t-x}, -e^{-t-x})$ in the Minkowski coordinates which is nothing but the original position that the $\phi_L(t,x)$ is inserted. Thus, the mirror operator $\mir(\phi_L(t,x))$ is a singular operator because it forcibly acts as $\phi_L(t,x) \in \mathcal{A}_L$ by an operator in $\mathcal{A}_R$.

In a rigorous sense, the local field $\phi_L(t,x)$ is not well-defined because it contains arbitrary large momentum $k\sim \infty$, and we should smear it. 
Let us consider the following operator smeared by the Gaussian distribution centering at $t_L=0, x_L=x_0$ with the width $\Delta_R$ (in the Rindler coordinates):
\begin{align}
    \phi_{\Delta_R}(x_0)&=\frac{1}{\sqrt{2\pi \Delta_R^2}}\int^{\infty}_{-\infty}\!dx\,e^{-\frac{(x-x_0)^2}{2\Delta_R^2}}\phi_{L}(t=0,x)
    \nn
    &=\int^{\infty}_0\frac{dk}{4\pi k}\left(e^{-\frac{\Delta_R^2}{2}k^2+ik x_0}a_k^L +e^{-\frac{\Delta_R^2}{2}k^2-ik x_0} a_k^{L\dagger} \right),
    \label{smear-psiL}
\end{align}
where we have used the e.o.m. of the right-moving free scalar: $\phi_{L}(t,x)=\phi_{L}(0,x-t)$.
The smeared operator effectively consists of the low-momentum modes $k \leq 1/\Delta_R$. 

On $\ket{TFD}$, the mirror operator of this smeared operator is given by
\begin{align}
   \mir(\phi_{\Delta_R})&= \frac{e^{\frac{\beta^2}{8\Delta_R^2}}}{\sqrt{2\pi \Delta_R^2}}\int^{\infty}_{-\infty}\!dx\,
   e^{-\frac{(x+x_0)^2-i\beta (x+x_0)}{2\Delta_R^2}}
   \phi_{R}(t=0, x)
   \nn
   &=e^{\frac{\beta^2}{8\Delta_R^2}}\int^{\infty}_0\frac{dk}{4\pi k}\left(e^{-\frac{\Delta_R^2}{2}\left(k-\frac{\beta}{2\Delta_R^2}\right)^2-i k x_0} a_k^{R} +e^{-\frac{\Delta_R^2}{2}\left(k+\frac{\beta}{2\Delta_R^2}\right)^2+i k x_0 }a_k^{R\dagger}\right).
\end{align}
From the first equation, we can find that the mirror operator is centered at $x_R=-x_0$ with the width $\Delta_R$ but the smearing function also contains a momentum-shifting factor $ e^{\frac{i\beta (x+x_0)}{2\Delta_R^2}}$.
From the second equation, we can explicitly see that the dominant momentum is $k\sim \frac{\beta}{2\Delta_R^2}$ for annihilation modes $a_k^{R}$, and these modes have exponentially large contributions due to the overall factor $e^{\frac{\beta^2}{8\Delta_R^2}}$. On the other hand, all the creation modes have the suppression factor $e^{-\frac{\Delta_R^2}{2}\left(k+\frac{\beta}{2\Delta_R^2}\right)^2}$ because the integration range is $0 \leq k \leq \infty$. Combining the overall factor $e^{\frac{\beta^2}{8\Delta_R^2}}$, only small momenta  $k\sim 0$ has $\mathcal{O}(1)$ contributions to the creation modes. 

The dominant momentum $k\sim \frac{\beta}{2\Delta_R^2}$ for annihilation modes is very large for well localized operator $\Delta_R \ll \beta$.
This is problematic if we think that our free theory is an effective field theory (EFT) with cutoff $k \lesssim \Lambda$.
The original smeared operator \eqref{smear-psiL} can be described by the EFT if the smearing width is larger than the cutoff length $\Delta_R \gtrsim 1/\Lambda^{-1}$. However, the mirror operator $\mir(\phi_{\Delta_R})$ is problematic if $\Delta_R \ll \beta$ because the dominant momentum can be greater than the cutoff as $k\sim \frac{\beta}{2\Delta_R^2} \gg \Lambda$. Then, the mirror operator description is not valid in the EFT. 

The smearing width $\Delta_R$ which we have considered is the coordinate width, and the physical width is 
$e^{-x_0}\Delta_R$ for $\Delta_R\ll 1$. 
Thus, if we considered a smeared field localized well in the region $X_0-\frac{\Delta_M}{2} \leq X \leq X_0+\frac{\Delta_M}{2}$ in the Minkowski coordinate, we should take $x_0 =-\log(-X_0)$ and $\Delta_R=e^{x_0}\Delta_M=\frac{\Delta_M}{(-X_0)}$ where we suppose $X_0$ is negative to consider the operator supported on the left wedge. 
It means that if the center of the smeared operator is near the Rindler horizon $X_0 \sim 0$, $\Delta_R$ can be $\mathcal{O}(1)$ even if $\Delta_M \ll 1$. Thus, in this case, the dominant momentum $k\sim \frac{\beta}{2\Delta_R^2}$ does not become large and the above problem of the mirror operator may be avoided.
Nevertheless, far from the horizon, the mirror map is still problematic.

We finally remark that the mirror map differs from the holographic subregion map considered in section~\ref{sec:comp}, i.e., the map from $\phi^G$ to $\phi^R$ in CFT. 
The mirror map does not preserve the Hermiticity of operators as can be understood from \eqref{mir-b}, and does not preserve the order of the multiplicity as \eqref{multi-mir}, while the holographic subregion map preserves them.


\section{On AdS-Rindler mode expansion}
\label{mode}

Here, we concentrate on the $d=2$ case for simplicity.
For the free scalar field theory in the three-dimensional AdS space, the bulk local operator $\phi(\tau, \rho, \theta)$ in the global patch can be expanded by the modes in the global AdS as
\begin{align}
\phi(\tau, \rho, \theta) =
\sum_{n,m} 
\left( 
a^{ \rm global \,\, \dagger}_{nm} \, e^{i \omega_{n m} \tau}  e^{-i m \theta}
+a^{\rm global}_{n m} e^{-i \omega_{n m} \tau}  e^{i m \theta}
\right)
\psi_{nm}^\text{bulk}(\rho)
\end{align}
where $n$ is a non-negative integer, $m$ is an integer,
$\omega_{nm}=2n+|m|+\Delta$
and $\psi_{nm}^\text{bulk}(\rho)$ are modes in $\rho$-direction. 
The explicit form is given by $\psi_{nm}^\text{bulk}(\rho)=\frac{1}{\mathcal{N}_{nm}} \sin^{|m|} (\rho) \cos^\Delta (\rho)\,
P_n^{|m|, \Delta-1} \left( \cos(2 \rho) \right)$, where  
$\mathcal{N}_{nm}$ is the numerical constant given in \cite{Fitzpatrick:2010zm} and $P_n^{|m|, \Delta-1}$ is the Jacobi polynomial.

We now take the half subregion (i.e. take $\theta_0=\pi/2$ in Appendix~\ref{app:AdSRindler}). 
In the right AdS-Rindler wedge,
the mode expansion of the bulk local operator
is given by
\begin{align}
\label{d=2_phi}
    \phi(t_R, \xi,\chi)=\int^{\infty}_{0}d\omega \int^{\infty}_{-\infty} d \lambda \frac{1}{\sqrt{2\pi}}
    \tilde{\psi}_{\omega, \lambda}(\xi) 
    \left[
    a_{\omega,\lambda} e^{-i\omega t_R +i \lambda \chi}+a^{\dagger}_{\omega,\lambda} e^{i\omega t_R -i \lambda \chi}
    \right]
\end{align}
where
\begin{align}
\label{psiR}
\tilde{\psi}_{\omega, \lambda}(\xi) 
=\frac{N_{\omega,\lambda}}{\Gamma(\nu+1)} \xi^{i\omega} (1+\xi^2)^{-\frac{i\omega}{2}-\frac{\Delta}{2}} 
    ~_2F_1\left(\frac{i\omega-i\lambda+\nu+1}{2} ,\frac{i\omega+i\lambda+\nu+1}{2}
  ;\nu+1;\frac{1}{1+\xi ^2}\right),
\end{align}
with $\nu=\Delta-\frac{d}{2}=\Delta-1$ and
\begin{align}
    N_{\omega,\lambda}=\frac{|\Gamma\left(\frac{i\omega-i\lambda+\nu+1}{2}\right)|\, |\Gamma\left(\frac{i\omega+i\lambda+\nu+1}{2}\right)|}{\sqrt{4\pi\omega} |\Gamma(i\omega)|}.
\label{nc}
\end{align}
The coordinate transformation of the global and AdS-Rindler coordinates are explicitly given by
\begin{align}
    \cos \rho =\frac{1}{\sqrt{1+\xi^2}\cosh \chi}, \,\,\, \tan \theta = \frac{\sqrt{1+\xi^2}}{\xi} \sinh \chi,
    \label{ct1}
\end{align}
on the $\tau=0$ ($t_R=0$) slice.

Let us consider the following smeared bulk local operator on $t_R=0, \xi=\xi_0, \chi=\chi_0$ in the AdS-Rindler wedge:
\begin{align}
\label{sblo}
    \phi^{\Lambda}(\xi_0,\chi_0)=
    \frac{1}{2 \pi \Lambda \Lambda'} 
    \int^{\infty}_{0} d\xi\int^{\infty}_{-\infty} d \chi 
     e^{-\frac{\Lambda^2}{2} (\xi-\xi_0)^2 -\frac{{\Lambda'}^2}{2} (\chi-\chi_0)^2}
     \phi(t_R=0, \xi,\chi).
\end{align}
Here, $\Lambda, \Lambda' \gg 1$, $\Lambda' = \mathcal{O} (\Lambda)$ and $\Lambda \xi_0 \gg 1$.\footnote{
With the last condition, the Gaussian factor is small on $\xi=0$. Instead of $\xi$, we can consider the Gaussian smearing for, for example, $\xi'=\ln(e^\xi-1)$, which takes $-\infty < \xi' < \infty$.
Such a modification is not important for the qualitative properties of $\phi^{\Lambda}(\xi_0,\chi_0)$ and the results below are unchanged. 
}
We will expand it by the global modes as 
$\phi^{\Lambda}(\xi_0,\chi_0)=\sum_{n,m} \beta^{n m} a_{n,m}^{ \rm global }+h.c.$
and evaluate the coefficient
\begin{align}
\label{beta1}
     \beta^{n m} =\frac{1}{2 \pi \Lambda \Lambda'} 
    \int^{\infty}_{0} d\xi\int^{\infty}_{-\infty} d \chi 
     e^{-\frac{\Lambda^2}{2} (\xi-\xi_0)^2 -\frac{{\Lambda'}^2}{2} (\chi-\chi_0)^2}
     e^{i m \theta} \psi_{nm}^\text{bulk}(\rho).
\end{align}
In order to do so, we need the large $n,m$ behavior of $\psi_{nm}^\text{bulk}(\rho)$.
The bulk e.o.m. is approximately given by
\begin{align}
     ( (\partial_\rho)^2+(2n +|m|)^2 - \frac{1}{\sin^2 \rho} m^2 ) \psi_{nm}^\text{bulk}(\rho) \simeq 0,
\end{align}
and the solution around $\rho=\rho_0$ is $\psi_{nm}^\text{bulk}(\rho_0 + \delta \rho) \simeq C_+ e^{i n \alpha \delta \rho} + C_- e^{-i n \alpha \delta \rho}$ with
\begin{align}
\alpha^2 = 4(1+a)- \frac{a^2}{\tan^2 \rho_0},      
\end{align}
where $C_\pm$ are $n$-independent constants and $a \equiv |m|/n$.
Here, $\rho_0=\rho(\xi_0,\chi_0)$ by the coordinate transformation \eqref{ct1}. 
Thus, we have $\psi_{nm}^\text{bulk}(\rho_0 + \delta \rho) \sim \cos (n \alpha \delta \rho +C)$ 
for $\alpha^2>0$ or exponential for 
$\alpha^2<0$.
This is consistent with the asymptotic behavior of the Jacobi polynomial given in
\cite{1360576122179052928, doi:10.1137/0522092},
where $\psi_{nm}^\text{bulk}(\rho) \sim n^{C(\rho,a)} \cos (A(\rho,a) n +B(\rho,a) )$
for $\alpha^2>0$, which can be written as $a<2 \frac{\sin \rho_0}{\cos^2 \rho_0} (1+\sin \rho_0)$,
and $\psi_{nm}^\text{bulk}(\rho) \sim n^{\tilde{C}(\rho,a)} e^{-\tilde{A}(\rho,a) n +\tilde{B}(\rho,a)}$
for $\alpha^2<0$.

Then,  $\beta^{nm}$ given in \eqref{beta1} for large $n, m$ is 
exponentially suppressed with the factor
$e^{-\tilde{A}(\rho_0,a) n}$ for $\alpha^2<0$. 
For $\alpha^2 >0$, 
by the Gaussian integration in \eqref{beta1} it has
the Gaussian damping factor  
$e^{-\frac{n^2}{2 \Lambda^2} (c_1)^2
-\frac{n^2}{2 {\Lambda'}^2} (c_2)^2}$,
where $c_1=
\left. \left(  \alpha \frac{\partial \rho}{\partial \xi} 
+a \frac{\partial \theta}{\partial \xi} 
\right) \right|_{\xi=\xi_0,\chi=\chi_0} $
and 
$c_2=
\left. \left(  \alpha \frac{\partial \rho}{\partial \chi} 
+a \frac{\partial \theta}{\partial \chi} 
\right) \right|_{\xi=\xi_0,\chi=\chi_0} $
for $n \gg \Lambda$.
Thus,
for generic $\xi_0,\chi_0$,
the smeared bulk local operator $\phi^{\Lambda}(\xi_0,\chi_0)$ in the AdS-Rindler patch
contains mostly low energy modes of 
$a_{n,m}^{ \rm global }$.
Here, low energy means $n,|m| =\mathcal{O}(\Lambda)$.

However, if the center $(\xi_0, \chi_0)$ of the smearing is near the AdS-Rindler horizon $\xi=0$, the smeared bulk local operator contains the high energy modes as we will see below.
Near $\xi=\xi_0, \chi=\chi_0$ with $\xi_0 \simeq 0$, the metric becomes approximately
\begin{align}
    ds^2 \simeq \xi_0^2 (- t_R^2 +d \zeta^2 +d \tilde{\chi}^2),
\end{align}
where
\begin{align}
    \zeta= \frac{\xi-\xi_0}{\xi_0}, \,\,\,\, \tilde{\chi} =\frac{\chi-\chi_0}{\xi_0}.
\end{align}
Under this coordinate system, 
the cut-offs for the momenta of $\zeta, \tilde\chi$
will be identified as the cut-off of the energy that conjugates to the Rindler time $t_R$ because the metric is the one of the Minkowski spacetime.
We will denote these as $\Lambda^R, \Lambda'^R$, which 
are given by $\Lambda^R=\xi_0 \Lambda,\, \Lambda'^R= \xi_0 \Lambda' $ from \eqref{sblo}.
In the global coordinates, where the Rindler horizon is at $\theta=\pm \pi/2$, the coordinate transformation is
\begin{align}
    \delta \rho \simeq \frac{1}{\cosh \chi_0} \xi_0 \tilde\chi, \,\,\,\,
    \delta \theta \simeq \frac{1}{\sinh \chi_0} \xi_0 \zeta.
\end{align}
Using the large $n,m$ approximation,
$e^{i \theta m} \psi_{nm}^\text{bulk}(\rho) \sim e^{i m \delta \theta } \cos (n \alpha \delta \rho +C) $,
we find that 
$\beta^{nm}$ is exponentially small if $n \gg \frac{1}{\xi_0}\Lambda'^R$ or $|m| \gg \frac{1}{\xi_0}\Lambda^R$.\footnote{
The asymptotic boundary $\xi \rightarrow \infty$ is also 
special. 
Near $\xi=\xi_0, \chi=\chi_0$ with $\xi_0 \simeq \infty$, the metric becomes approximately $ds^2 \simeq \xi_0^2 (- t_R^2 +d \zeta^2 +d \chi^2)$
where $ \zeta= -\frac{\xi-\xi_0}{(\xi_0)^2}$.
The cut-off for the momenta of $\zeta$
is given by $\Lambda^R=(\xi_0)^2 \Lambda $.
With $\delta \rho \simeq - \frac{1}{\sqrt{1+\cosh^2 \chi}} \frac{1}{(\xi_0)^2} \delta \xi$ and $\tan \theta \simeq \sinh \chi$,
we find that $\beta^{nm}$ is exponentially small if $n \gg \Lambda^R$ or $|m| \gg \Lambda'$.
This is the same as for a generic $\xi_0, \chi_0$.
}

Thus, the smeared bulk local operator at $\xi=\xi_0$ in the AdS-Rindler patch
with the energy cut-off $\mathcal{O} (\Lambda^R)$ contains 
the modes in the global coordinate with the energy of $\mathcal{O}(\frac{1}{\xi_0}\Lambda^R)$.
Because the finite $N$ CFT has many  but finite degrees of freedom determined by $N$, there is an upper bound for the energy of the modes 
which is expected to be the Planck energy $\Lambda_\text{Planck}$.
If $\xi_0 \leq 1/\Lambda_\text{Planck}$,  because $\Lambda^R \gg 1$,
the smeared bulk local operator in the AdS-Rindler patch
consists of the trans-Planckian energy modes of global coordinates
and then it cannot be well-defined.
Note that the surface $\xi_0 =1/\Lambda_\text{Planck}$ is the 
stretched horizon, which is away from the horizon by the Planck length because the metric becomes $ds^2 \simeq -\xi_0  t_R^2 +d \xi^2 +d \chi^2$ near the horizon.
This is regarded as the brick wall argument \cite{tHooft:1984kcu, Iizuka:2013kma} without the artificial Dirichlet boundary condition for the AdS-Rindler case.

\section{Bulk coherent states}
\label{App:coherent}
Here we consider a free massive scalar field in $(d+1)$-dimensional AdS space (global patch). 
Let $f_{nlm}(x)$ be the positive frequency solutions of the EoM, 
\begin{align}
    f_{nlm}(x)=e^{-i \omega_{nl}\tau}\psi_{nl}(\rho)Y_{lm}(\Omega)
\end{align}
with 
\begin{align}
   &\omega_{nl}=2 n + l +\Delta,\quad \Delta=\frac{d}{2}+\sqrt{m^2+\frac{d^2}{4}}= \frac{d}{2}+\nu.
\end{align}
$\psi_{nl}(\rho)$ represents the modes for $\rho$-direction whose explicit form is not important here. 
We introduce the Klein-Gordon inner product in the global AdS as
\begin{align}
    (f_1,f_2):=
    i\int dV
   (f_1^\ast \partial_\tau f_2-(\partial_\tau f_1^\ast)f_2), 
   \label{def:KG_ip}
\end{align}
where $\int dV = \int^{\frac{\pi}{2}}_0 d\rho \int d\Omega_{d-1} 
    (\tan \rho)^{d-1}$. 
Using the inner product, we normalize the modes as $ (f_{nlm},f_{n'l'm'})=\delta_{nn'}\delta_{ll'}\delta_{mm'}$.

We can expand the real scalar field by these modes, 
\begin{align}
\phi(x)=\sum_{n,l,m}\left(a_{nlm}f_{nlm}(x)+a_{nlm}^\dagger f_{nlm}^\ast(x)\right).
\end{align}
In our normalization, we have the commutation relation
\begin{align}
    [a_{nlm}, a_{n'l'm'}^\dagger]=\delta_{nn'}\delta_{ll'}\delta_{mm'}.
\end{align}


Let us consider an arbitrary classical configuration $\phi_{cl}(x)$ that satisfies the classical EoM. 
This solution can be expanded by the modes $f_{nlm}$ as
\begin{align}
\label{phi_cl_exp}
   \phi_{cl}(x)= \sum_{n,l,m}\left(B_{nlm}f_{nlm}(x)+B_{nlm}^\ast f_{nlm}^\ast(x)\right)
\end{align}
with the coefficients $B_{nlm}$ given by the Klein-Gordon inner product
\begin{align}
    B_{nlm}=(f_{nlm}, \phi_{cl}).
    \label{Bnlm}
\end{align}
We then introduce the following anti-Hermitian operator
\begin{align}
    B=\sum_{n,l,m}\left(-B_{nlm}^\ast a_{nlm}+B_{nlm} a_{nlm}^\dagger\right).
    \label{def:B}
\end{align}
It satisfies 
\begin{align}
    [\phi(x),B]=\phi_{cl}(x),
\end{align}
where the right-hand side represents c-number $\phi_{cl}(x)$ times an identity operator. 
Thus, the state 
\begin{align}
    \ket{\Psi}=e^{B}\ket{0}
    \label{bulk_coherent_st}
\end{align}
is the coherent state corresponding to the classical configuration $\phi_{cl}(x)$ as
\begin{align}
    \bra{\Psi}\phi(x)\ket{\Psi}=\phi_{cl}(x)
\end{align}
where we have used that the vacuum 1-pt function $\bra{0}\phi\ket{0}$ vanishes.

Note also that the operator $B$ can be written using the Klein-Gordon inner product as
\begin{align}
   B= -(\phi_{cl},\phi)_{\tau=0}=-i\int_{\tau=0} dV
   (\phi_{cl} \partial_\tau \phi-(\partial_\tau \phi_{cl})\phi),
\end{align}
where note that $\phi$ is an operator but $\phi_{cl}$ is a c-number function.
By decomposing the integration region in the inner product, i.e. $\tau=0$ slice, into $M_A$ and $M_{\bar{A}}$, $B$ is decomposed as 
\begin{align}
    B=B_{M_A}+B_{M_{\bar{A}}}
\end{align}
with
\begin{align}
\label{eq:BMA}
    B_{M_A}=-i\int_{M_A} dV
   (\phi_{cl} \partial_\tau \phi-(\partial_\tau \phi_{cl})\phi),
   \quad  
   B_{M_{\bar{A}}}=-i\int_{M_{\bar{A}}} dV
   (\phi_{cl} \partial_\tau \phi-(\partial_\tau \phi_{cl})\phi).
\end{align}
Then, the state $\ket{\Psi}$ can be written as
\begin{align}
\label{psi-eBMA-eBbMA}
\ket{\Psi}=e^{B_{M_A}}e^{B_{M_{\bar{A}}}}\ket{0}
\end{align}
Note that $e^{B_{M_A}}, e^{B_{M_{\bar{A}}}}$ are unitary.

Since $\ket{\Psi}$ is a coherent state, we have
\begin{align}
  \bra{\Psi}\phi(x) \phi(x')\ket{\Psi}- \bra{0}\phi(x) \phi(x')\ket{0}=\phi_{cl}(x)\phi_{cl}(x').
\end{align}
The change in the expectation value of the energy-momentum tensor is
\begin{align}
\bra{\Psi}T_{MN}^\text{bulk}\ket{\Psi}- \bra{0}T_{MN}^\text{bulk}\ket{0}
    =T_{MN}^\text{bulk}[\phi_{cl}],
    \label{bulkTMN}
\end{align}
where $T_{MN}^\text{bulk}[\phi_{cl}]$ is the energy-momentum tensor for the classical configuration $\phi_{cl}$. The right-hand side of \eqref{bulkTMN} is UV-finite.

Here we explain again the argument of an incompatibility between the FLM proposal and the entanglement wedge reconstruction. The discussion is also based on the factorization of the bulk Hilbert space as remarked in the beginning of subsection~\ref{subsec:contra}.
We compare the entanglement entropy of the excited state corresponding to the bulk state $\ket{\Psi}=e^B \ket{0}$ to the one of the vacuum $\ket{0}$. 
The situation is similar to the setup in sec~4.1 in \cite{Jafferis:2015del}.
Supposing that exication is small as $\mathcal{O}(N^0)$, the FLM proposal states that the difference of the CFT entanglement entropy between $\ket{\Psi}$ and $\ket{0}$ satisfies
\begin{align}
\label{FLMapp}
    \delta S_A^\text{CFT}=\frac{\delta A_\Sigma}{4 G_N}+\delta S_{M_A}^\text{bulk}
\end{align}
at the leading order of the $1/N$ expansion. Here, the `leading' means the first leading correction, $\mathcal{O}(N^0)=\mathcal{O}(G_N^{0})$ terms, in the $1/N$ expansion because $\mathcal{O}(N^2)=\mathcal{O}(G_N^{-1})$ terms are canceled out.

First, let us consider the last term in the right-hand side of \eqref{FLMapp}.
We have \eqref{psi-eBMA-eBbMA} and $e^{B_{M_A}}, e^{B_{M_{\bar{A}}}}$ are unitary operators acting only on $\cH_{M_A}$ and $\cH_{\bar{A}}$ respectively. 
Thus, the last term $\delta S_{M_A}^\text{bulk}$ vanishes because any local unitary transformation does not change the entanglement entropy.

Next consider the first term $\frac{\delta A_\Sigma}{4 G_N}$.
The excited state $\ket{\Psi}$ of a bulk matter field has $\mathcal{O}(G_N^0)$  bulk stress tensor given by \eqref{bulkTMN}.
Then, the leading back-reaction to the metric is obtained by solving the linearized Einstein equation with the source corresponding to this stress tensor. Schematically, EoM takes the form
$E_g(\delta g)_{MN}=8\pi G_N T_{MN}^\text{bulk}$, where $E_g(\delta g)_{MN}$ represents the (left-hand side of) linearized Einstein eq.
Since $T_{MN}^\text{bulk}$ is $\mathcal{O}(G_N^0)$, the change of metric $\delta g_{MN}$ is $\mathcal{O}(G_N)$, and thus the change of area $\delta A$ is $\mathcal{O}(G_N)$.
Then $\frac{\delta A_\Sigma}{4 G_N}$ can be non-zero as follows. 
$T_{MN}^\text{bulk}$ depends on $\phi_{cl}$, and the choice of $\phi_{cl}$ is arbitrary. For any choice of $\phi_{cl}$, we can construct the corresponding coherent state as explained above. 
(For any $\phi_{cl}$ whose mode expansion is given by \eqref{phi_cl_exp}, we can construct $\ket{\Psi}=e^B \ket{0}$ with $B$ given by \eqref{def:B}.)
Since configuration $\phi_{cl}$ is arbitrary, we can take it so that the back-reaction $\delta g_{MN}$ at the RT surface $\Sigma$ does not vanish. 
$\frac{\delta A_\Sigma}{4 G_N}$ can be non-zero, and the right-hand side of \eqref{FLMapp} does not have to vanish. 

However, the left-hand side of \eqref{FLMapp} must vanish if the (strong) entanglement wedge reconstruction holds.
Since $B_{M_A}$ in \eqref{eq:BMA} is written only by operators $\phi, \partial_\tau \phi$ on $M_A$ and thus it should be represented by CFT operators on $A$, if we adopt the (strong) entanglement wedge reconstruction. 
Similarly, 
$B_{M_{\bar{A}}}$ should be represented by CFT operators on $\bar{A}$.
$B=B_{M_A}+B_{M_{\bar{A}}}$ where $B_{M_A}, B_{M_{\bar{A}}}$ are operators supported only in $M_A, M_{\bar{A}}$ respectively.
Thus, the operators $e^{B_{M_A}}$ and $e^{B_{M_A}}$ are the unitary operators on $A$ and $\bar{A}$. 
Therefore, the entanglement entropy on $A$ for $\ket{\Psi}$ is the same as for $\ket{0}$, and hence the left-hand side of \eqref{FLMapp} vanishes. 
It contradicts with the fact that the right-hand side can be non-zero.
Of course, if we do not assume the entanglement wedge reconstruction, we do not encounter this contradiction.

\newpage 

\bibliographystyle{utphys}
\bibliography{ref-AdSCFT}
\end{document}